\documentclass[12pt,eqsecnum,preprint,flushrt]{aastex}
\slugcomment{Submitted to ApJ}
\usepackage{graphicx}
\usepackage{vmargin}
\usepackage{amssymb}

\def\pd#1#2{\frac{\partial #1}{\partial #2}}
\def\od#1#2{\frac{d #1}{d #2}}
\def\pdd#1#2{\frac{\partial^2 #1}{\partial #2^2}}

\newcommand{\Del}{\mathbf{\nabla}}

\def\ahalf{{\frac{1}{2}}}
\def\oneover#1{{\frac{1}{#1}}}
\def\prn#1{{\left(#1\right)}}
\def\brk#1{{\left[#1\right]}}
\def\brc#1{{\left\{#1\right\}}}
\def\abs#1{{\left|#1\right|}}
\newcommand{\sech}{\text{sech}}

\newcommand{\zhat}{\mathbf{\hat z}}

\newcommand{\text}[1]{\rm{#1}}
\newcommand{\eqref}[1]{(\ref{#1})}
\newcommand{\iint}{\int\!\!\!\int}

\title{X-winds in Action}
\author{Mike J. Cai$^1$, Hsien Shang$^1$, Hsiao-Hsuan Lin$^{2}$, Frank H. Shu$^3$
} \affil{$^1$Academia Sinica, Institute of Astronomy
and Astrophysics, Taiwan\\
$^2$Department of Physics, University of Southern California\\
$^3$Department of
 Physics, University of California, San Diego}
\email{mike@asiaa.sinica.edu.tw}

\begin{abstract}
The interaction of accretion disks with the magnetospheres of young
stars can produce X-winds and funnel flows. With the assumption of
axial symmetry and steady state flow, the problem can be formulated
in terms of quantities that are conserved along streamlines, such as
the Bernoulli integral (BI), plus a partial differential equation
(PDE), called the Grad-Shafranov equation (GSE), that governs the
distribution of streamlines in the meridional plane.  The GSE plus
BI yields a PDE of mixed type, elliptic before critical surfaces
where the flow speed equals certain characteristic wave speeds are
crossed and hyperbolic afterward.  The computational difficulties
are exacerbated by the locations of the critical surfaces not being
known in advance.  To overcome these obstacles, we consider a
variational principle by which the GSE can be attacked by
extremizing an action integral, with all other conserved quantities
of the problem explicitly included as part of the overall
formulation.   To simplify actual applications we adopt the cold
limit of a negligibly small ratio of the sound speed to the speed of
Keplerian rotation in the disk where the X-wind is launched.  We
also ignore the obstructing effects of any magnetic fields that
might thread a disk approximated to be infinitesimally thin. We then
introduce trial functions with adjustable coefficients to minimize
the variations that give the GSE.  We tabulate the resulting
coefficients so that other workers can have analytic forms to
reconstruct X-wind solutions for various astronomical,
cosmochemical, and meteoritical applications.
\end{abstract}

\begin{document}
\maketitle
\section{Introduction}

Accretion, disks, and jets are ubiquitous in astrophysics (see,
e.g., Blandford \& Rees 1992).  A consensus has been reached that an
extra needed ingredient to obtain outflow from inflow is the
presence of strong magnetic fields that thread a disk conventionally
assumed to be rotating at Keplerian speeds about a central
gravitating object, taken in this paper to be a newly born star.
Differences come in ascribing the origin of the magnetic fields in
the disk itself or in the central star (K\"onigl \& Pudritz 2000,
Shu et al. 2000).

Disk winds have been extensively studied, both analytically via the
assumptions of self-similarity in 2-D space for axisymmetric,
time-independent flows (e.g., Blandford \& Payne 1982;  Contopoulos
\& Lovelace 1994) or by taking advantage of arbitrary variations of
the gas pressure (e.g., Tsinganos \& Trussoni 1991) or by studying
the asymptotic properties of the collimation (Heyvaerts \& Norman
1997); and numerically by finite-element methods attacking the
axisymmetric, time-independent, Grad-Shafranov equation (e.g., in
the relativistic regime by Camenzind 1987) or by finite-difference
treatments of the time-dependent equations of ideal MHD in 2- and
3-D (e.g., Uchida \& Shibata 1986, Pudritz et al. 2006). For a
review of these types of calculations, see Ferreira (2004).

The most highly developed semi-analytic theory for the second
viewpoint is called X-wind theory, in which fast jets arising in
young stellar objects (YSOs) owe their existence to the interaction
of the accretion disk with the magnetosphere of the central star.
The interaction of accretion disks with strongly magnetized central
stars has also been studied numerically (e.g., Goodson, Bohm, \&
Winglee 1999; Long et al. 2005, Ustyugova et al. 2006).   Although
both funnel flows and X-like winds have been found, they have yet to
appear simultaneously in numerical simulations, probably because the
numerical calculations have not yet proceeded to steady state where
the condition of disk-locking applies (Shu et al. 1994).  Pure
X-wind theory assumes for simplicity that the disk itself is
unmagnetized, in fact, all that is needed for the theory to work is
for open field lines to be concentrated in a narrow annulus near the
inner edge of an accretion disk.

Recently, Bacciotti et al. (2002) and Coffey et al. (2004)
identified jet rotation in four T Tau systems, DG Tau, TH 28, RW
Aur, and LkH$\alpha$ 321, of an amount too large to be compatible
with X-winds, but consistent with launching from disks at radii of
0.5-2 AU.  Later, Cabrit et al. (2006) showed from mm-wave radio
measurements that the disk rotation in RW Aur is actually in the
opposite sense to that deduced for the jet from optical lines.
Moreover, Pety et al. (2006) find that HH 30, which is observed
nearly edge-on and therefore should have had the clearest signature
for jet rotation, showed no evidence for outflow rotation at mm
wavelengths, a conclusion reinforced by optical and ultraviolet
observations of the HH 30 jet by Coffey et al. (2007). While the
positive results remain for the three other systems, the case of HH
30, where longitudinal velocities occur in the direction transverse
to the line of sight, suggests that the slight line asymmetries in
the other cases may be more associated with unequal jumps in the
velocity of shocked, high-speed jets, than to the rotation of
collimated outflows.

In contrast, no one has proposed any explanation other than X-winds
for the correlated inflow-outflow signatures seen in SU Aur by
Giampapa et al. (1993) and Johns \& Basri (1995).  Apart from SZ 68
(Johns-Krull \& Hatzes 1997), we are unaware of any other T Tau star
that shows a tilted-dipole magnetic-field geometry, and it could be
that the dipole component is small on the surface of most T Tau
stars (Johns-Krull 2007).  Fortunately, Mohanty \& Shu (2007) show
that while funnel flows are sensitive to the detailed assumptions
made concerning multi-pole structure on the surfaces of the central
stars, the properties of the X-wind depend mostly only on the amount
of trapped flux in the X-region (see also the observational evidence
relating to this point collected by Johns-Krull \& Gafford 2002).

Recent calculations show that YSOs are unlikely to lose enough
magnetic flux in the process of gravitational collapse to make the
level of magnetization ignorable in the resultant circumstellar
accretion disks (Galli et al. 2006; Shu et al. 2006, 2007).  Indeed,
the disks are sufficiently magnetized in many cases that, in
quasi-steady state, they rotate at sub-Keplerian rates until the the
inner disk-edge is reached.  Thus, there are open questions of how
much of the trapped flux near the inner edge is to be attributed to
the central star versus the disk, and how such disks reacquire
near-Keplerian rates of rotation at their inner edges. We ignore
these complications in the present study of the X-wind phenomenon,
but we note that the methods introduced here are easily modified to
attack the more complex problem when the accretion disk interacting
with a stellar magnetosphere is itself strongly magnetized.

The original X-wind model supposed the outflow to occur from the
equator of a magnetized star spun to breakup by a presence of an
accretion disk that abutted its surface (Shu et al. 1988). Later, in
order to accommodate the slow rotators, such as the classical T
Tauri stars which are only rotating at one tenth of breakup (Vogel
\& Kuhi 1981, Bouvier et al. 1991, Edwards et al. 1993), Shu et al.
(1994a) generalized the X-wind picture to include the case of
relatively low accretion when the magnetosphere of the star would
truncate the accretion disk at an inner edge before the disk reached
the stellar surface (typically a circle of radius 0.2 on the scale
of Fig.1, where the disk's inner edge is taken to be at $\varpi =
1$).  In a quasi-steady-state where most of the mass of the central
star is built up by disk accretion, the magnetic coupling between
the star and the disk regulates the star to corotate at the
Keplerian frequency at the truncation radius. For a protostar with
magnetic dipole moment $\mu_*$, mass $M_*$, mass accretion rate
$\dot M_D$, Ostriker \& Shu (1995) estimate this radius to be
\begin{equation}
  R_X = \Phi_{\text{dx}}^{-4/7} \prn{\frac{\mu_*^4}{GM_*\dot
  M_D^2}}^{1/7},
\end{equation}
where $\Phi_{\text{dx}}$ is an order unity dimensionless number that
parameterizes the amount of stellar magnetic flux that is trapped in
the disk.  Inside this radius, matter is channeled to the star via a
funnel flow.  The excess angular momentum of the
accreting material is deposited in the magnetic field in the form of
Maxwell torque, and then transported back to the disk.  The gain of
angular momentum and approximate field freezing would try to move
the footpoint of the funnel-flow field lines outward.

Exterior to the truncation radius $R_X$, the equatorial inward drift
in the accretion disk creates an angle between the stellar
magnetic-field lines and the disk normal.  If approximate field
freezing holds as the accretion proceeds, a fraction of the field
lines will develop an angle larger than $30^\circ$, when matter
frozen to this flux tube becomes unstable to magnetocentrifugal
fling (Blandford \& Payne 1982).  These field lines are thus
responsible for driving a magnetohydrodynamic (MHD) wind from the
disk. Since the wind removes angular momentum from the disk, the
footpoints of those field lines in the disk will try to migrate
inward. The radially inward press of the footpoints of the wind
field lines footpoints and the radially outward press of the
footpoints of the funnel-flow field lines create a magnetic
X-configuration that distinguishes the model from similar variants
in the literature (the historical choice of the name from the
X-point of the equivalent gravitational potential in the co-rotating
frame is common to many models). In quasi-steady state where radial
advection into the X is balanced by the resistive diffusion of field
lines out of the X, Shu et al. (1994b) estimated that enough stellar
flux could be trapped in a small X-region near the inner disk edge
to have large dynamical effects, namely, the truncation of the disk
by a funnel flow out of the disk plane accompanied by an X-wind that
carries away most of the excess angular momentum transported into
the X-region.

Apart from the original numerical estimates, there are reasons to
suppose that if turbulent resistivity is the source of the diffusion
across magnetic field lines (Shu et al. 2007), then the fractional
size of the X-region in units of $R_X$ is given by the ratio of
sound speed at the surface of the disk where the X-wind is launched
to the local Keplerian speed at $R_X$.  For the inner disk of a
classical T Tauri star, the thermal sound speed is $a\sim 5$ km/s
while the Keplerian speed at $R_X$ is $v_K \sim 100$ km/s (Najita et
al. 2007); thus, the ratio $\epsilon$ is a small number $\sim 0.05$.
In an asymptotic analysis where $\epsilon$ is taken to $\rightarrow
0$, the X-wind tied to the trapped field lines in the X-region would
emerge from virtually a single point in the meridional plane with a
fan-like geometry. Seen by an observer rotating at the Keplertian
angular frequency of $R_X$, gas flows along streamlines that
coincide with field lines if field freezing is assumed, and both
patterns of streamlines and field lines remain stationary in the
corotating frame.

\begin{figure}[ht]
\begin{center}
\includegraphics[width = 4in, angle = 0]{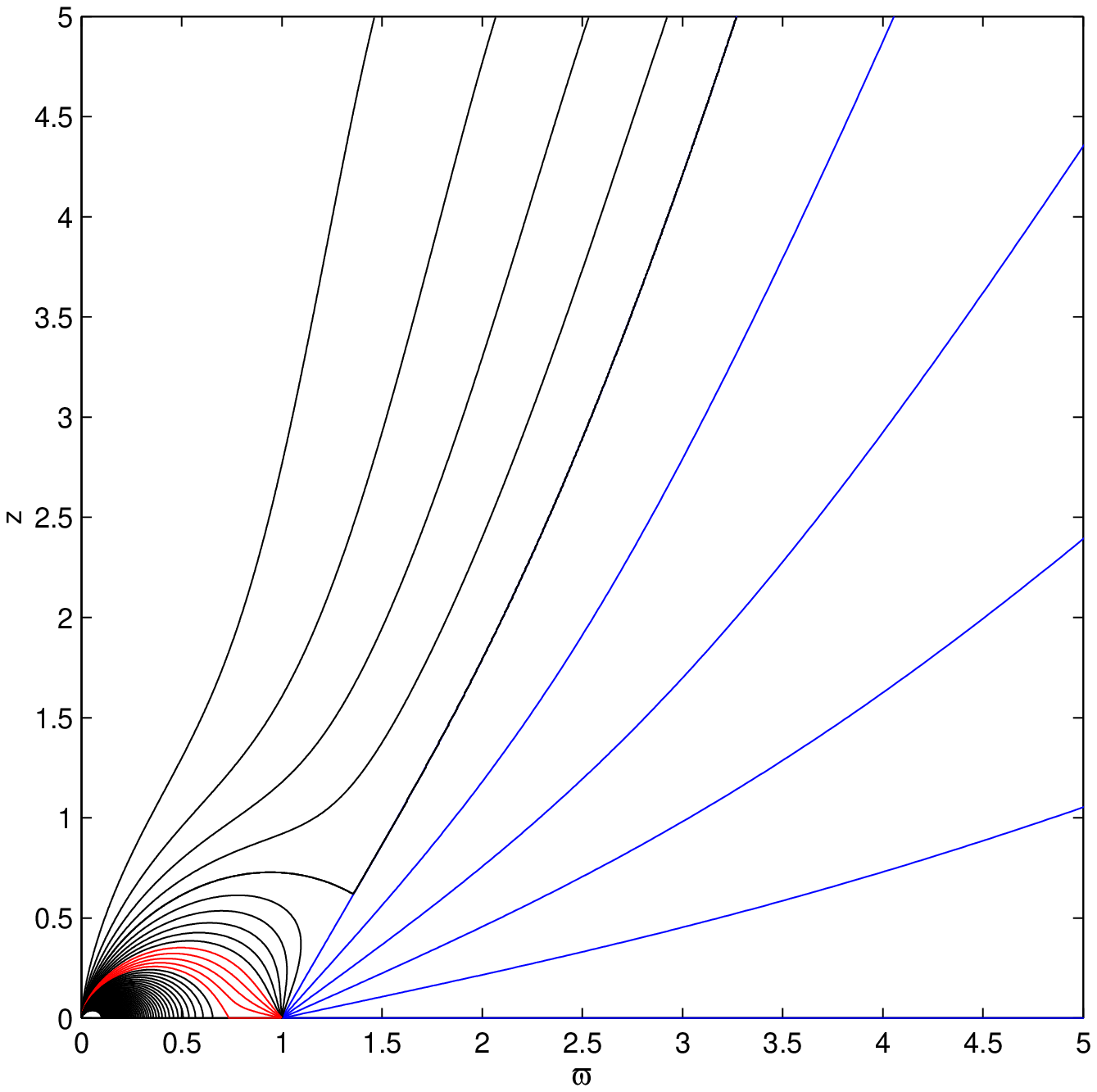}
\end{center}
\caption{Funnel flow (red curves), X-wind streamlines (blue curves),
and field lines dead to magnetocentrifugal fling (black curves)
according to Ostriker \& Shu (1995) and Shang et al. (1998).  The
magnetic field near the origin (center of YSO) is modeled as a
magnetic dipole, and all of the field lines contained in the X-wind
have their counterparts (with reversed directions) in opened stellar
field lines that lie inside a hollow cone dead to flow surrounding
the $z$-axis. Exact pressure balance across the sheet current that
divides the X-wind and dead field lines holds near and far from the
Y-point at $\varpi \approx 1.3$, $z\approx 0.7$, but this balance is
only approximate at intermediate distances, which accounts for why
the tilted upside-down Y of the separatrix does not have the equal
angles of 120$^\circ$ that would characterize an exact
Y-configuration appropriate for coronal conditions. In fact, the
field lines in red and black are computed as if they were vacuum
magnetic fields; only the portion depected in blue are attacked via
the solution of the Grad-Shafranov equation or its
variational-principle analog discussed in the present paper.  The
neutral line separating the black and blue field lines is replaced
by a separatrix of prescribed locus satisfying the approximate
pressure balance as described above. (see \S 4).}  The implied
poloidal and toroidal current flows are discussed in Ostriker \& Shu
(1995) and Shu et al. (1995); in particular, there is no current
flow along the $z$-axis which is taken to be a region of vacuum
longitudinal field. \label{funnel-wind}
\end{figure}

Viewed in this fashion, the overall problem can be broken into
smaller pieces and tackled separately. Using a formulation with a
precedent in the work of Lovelace et al. (1986), Shu et al. (1988,
1994b) wrote down the mathematical equations that describe a steady
state axisymmetric flow in the corotating frame (Grad \& Rubin 1958,
Shafranov 1966).  By the method of matched asymptotic expansions,
Shu et al. demonstrated the existence of an inner solution in the
X-region where the flow makes a sonic transition, and an outer
solution where the sound speed is formally taken to zero and the
X-region shrinks down to a point.  Najita \& Shu (1994) computed
numerically the portion of the X-wind in which the fluid velocity is
sub-Alfv\'enic, and the governing equation is elliptic.  Ostriker \&
Shu (1995) solved the problem of the funnel flow and the field
configuration in the dead zone (in which the field lines do not
depart sufficiently from disk normal to load any matter) in an
approximation that treated the accretion flow onto the star as
highly sub-Alfv\'enic. Shu et al. (1995) constructed asymptotic
solutions that describe the logarithmically slow, far-wind
collimation into jets.  The free boundaries between various parts of
the problem (funnel flow, dead zone, and X-wind) were determined by
pressure balance on either side.

In this paper, we wish to address the X-wind part of the overall
problem.  In order for the X-wind to accelerate from rest to
supersonic speeds, it must smoothly pass three surfaces on which the
flow velocity is equal to the slow MHD, Alfv\'en, and fast MHD
velocity, respectively (Heinemann \& Olbert 1978, Sakurai 1985).
These critical surfaces manifest themselves as singularities in the
governing equation (see Weber \& Davis 1967), and thus need to be
handled analytically.  In an axisymmetric problem, if the shapes of
the streamlines are known in the meridional plane, the conserved
quantities of the problem (mass to flux loading, angular momentum
flux including that carried in the Maxwell stress, and Bernoulli's
integral along a streamline) suffice to give a completely analytic
solution, including the locations and conditions required to cross
the critical surfaces smoothly. Unfortunately, the streamline
distribution in the meridional plane is not known a priori but must
be obtained, in principle, from a solution of the Grad-Shafranov
equation (GSE). The spatial location of the critical surfaces are
part of the overall solution of the GSE; indeed, they characterize
the regions where this PDE is elliptic or hyperbolic. The mixed
character of the GSE makes a direct numerical attack extremely
difficult when self-similarity does not apply, perhaps the hardest
problem in the mathematical theory of nonlinear PDEs of second-order
(see Garabedian 1986). The current work sidesteps the mathematical
solution of the GSE as a nonlinear PDE of second order, and
approaches it instead as a much more amenable problem in variational
calculus.

The rest of this paper is organized as following. In \S 2 we review
the basic formulation in terms of a stream function and an Alfv\'en
discriminant that yields the partial-differential equations -- the
so-called Grad-Shafranov and Bernoulli equations -- that govern a
steady, axisymmetric, X-wind flow.  In \S 3 we write down an action
whose variations with respect the stream function $\psi(\varpi,z)$
and the Alfv\'en discriminant ${\cal A}(\varpi,z)$ yield,
respectively, the Grad-Shafranov equation and the Bernoulli equation
when the action reaches an extremum. We also perform a
transformation where we replace the vertical coordinate $z$ in
cylindrical coordinates $(\varpi, \varphi, z)$ by $\psi$.  In \S 4,
we show how to incorporate boundary conditions into the problem, as
well as how to take advantage of the fact that analytic forms are
known for the solutions in the near-neighborhood of the X-point and
in the asymptotic regime far from the X-point (Shu et al. 1994b,
1995).  In \S 5, we outline a practical implementation of the
principle of extremal action, making use of only variations of
$\psi$ -- or, more precisely, of $z(\varpi,\psi)$ in our actual
working space -- as the substitute to attacking the Grad-Shafranov
equation, while we solve Bernoulli's equation directly for reasons
that are expounded upon in this section.  In \S 6, we present
numerical results for three specific cases of mass-loading onto wind
flux-tubes, finding good agreement with previous approximate
solutions obtained by Shang (1998) that have been used for many
different astrophysical applications (e.g., Shang et al. 1998, 2002,
2004).  In \S 7, we summarize the recipes needed to convert the
numerical solutions of \S 6 into practical dimensional models.  We
then offer our conclusions and suggestions for needed future
research.

\section{Basic Equations} \label{Basic Equations}
\newcommand{\Veff}{V_{\text{eff}}}
\newcommand{\A}{\mathcal{A}}
From the fundamental parameters of the problem, we may construct
units of length, time, and density as $R_X$, $\Omega_X^{-1}$, and
$\dot M_w/4\pi R_X^3\Omega_X$, respectively.  By assuming
axisymmetry and stationarity in a frame that is rotating with
angular velocity $\Omega_X$, we may write down the dimensionless
governing equations in the above units.
\begin{mathletters}
\begin{eqnarray}
  &&\Del \cdot (\rho \mathbf{u}) = 0, \label{continuity_eqn}\\
  &&\Del \prn{\ahalf \abs{\mathbf{u}}^2} + (2 \mathbf{e}_z + \Del
  \times \mathbf{u}) \times \mathbf{u} = -\frac{\epsilon}{\rho} \Del \rho
  -\Del \Veff + \oneover{\rho} (\Del \times \mathbf{B}) \times \mathbf{B},
  \label{Euler_eqn}\\
  &&\mathbf{B} \times \mathbf{u} = 0,\label{field_freezing}\\
  &&\Del \cdot \mathbf{B} = 0, \label{monopole}
\end{eqnarray}
\end{mathletters}
where $\epsilon \equiv a/R_X \Omega_X$ is the sound speed measured
in units of Keplerian velocity at the X-point, and is assumed to be
a small parameter of the problem. The effective potential in the
corotating frame is defined as
\begin{equation}
  \Veff = -\frac{1}{\sqrt{\varpi^2 + z^2}} -\ahalf \varpi^2 +
  \frac{3}{2}.
  \label{Veff}
\end{equation}
Here we have added a constant term to the effective potential so
that its numerical value is zero at the X-point.

\subsection{Constants of Motion}
\newcommand{\oneoverrho}{\beta^2 - \varpi^2\A}
The continuity equation \eqref{continuity_eqn} is satisfied
identically if we define the poloidal velocity through a stream
function (Shu et al. 1988, 1994a):
\begin{equation}
  \rho u_\varpi \equiv \oneover{\varpi} \pd{\psi}{z}, \quad \rho
  u_z \equiv - \oneover{\varpi} \pd{\psi}{\varpi}.
\label{psidef}
\end{equation}
For steady state axisymmetric flow in the corotation frame, the
field freezing condition \eqref{field_freezing} demands that the
magnetic field and the velocity are related by (see, e.g., Mestel 1968)
\begin{equation}
  \mathbf{B} = \beta \rho \mathbf{u}.
\label{betadef}
\end{equation}
With this identification, the continuity equation
\eqref{continuity_eqn} and the absence of magnetic monopoles
\eqref{monopole} imply $\mathbf{u} \cdot \Del \beta = 0$.  In terms
of the stream function, this means $\beta$ is constant along each
streamline, or $\beta = \beta(\psi)$.

The Euler equation describes momentum and energy balance in three spatial
dimensions.  If we take the component along the fluid velocity by
taking the inner product of \eqref{Euler_eqn} with $\mathbf{u}$, we
obtain the Bernoulli's equation (BE) along streamlines
\begin{equation}
  \mathbf{u} \cdot \Del H = 0 \; \text{where} \; H \equiv \ahalf
  \abs{\mathbf{u}}^2 + \epsilon^2 \ln \rho + \Veff.
\end{equation}
In other words, $H = H(\psi)$, and the energy per unit mass of an
isothermal gas, including its specific enthalpy, is conserved along
a streamline in the corotating frame where the flow occurs parallel
to $\bf B$.  Similarly, if we take the toroidal component of the
Euler equation \eqref{Euler_eqn}, we obtain a third conserved
quantity along stream lines, the angular momentum of the gas
allowing for that part carried away by the Maxwell torque of the
field:
\begin{equation}
  J \equiv \varpi^2 + \varpi(1-\beta^2 \rho) u_\varphi = J(\psi).
\label{Jdef}
\end{equation}
As we shall see, the determination of the conserved quantities,
$H(\psi)$and $J(\psi)$, is achieved by demanding that the X-wind
crosses the slow MHD and fast MHD surfaces smoothly.  The loading of
mass onto flux, which is governed by $\beta(\psi)$ is freely
specifiable within certain limits to be detailed below.

The last component of the Euler equation describes momentum balance
in the direction perpendicular to the poloidal field lines. It is
the famous Grad-Shafranov equation (Heinemann \& Olbert 1978, Sakurai 1985):
\begin{equation}
  \Del \cdot (\A \Del \psi) + \oneover{\A}\prn{\frac{J}{\varpi^2}
  -1}\frac{J'}{\varpi^2}
  + \frac{{\beta^2}'\brc{\Veff + \epsilon^2 \ln \brk{\epsilon^2
  h/(\oneoverrho)}}}{(\oneoverrho)^2}
  - \frac{\epsilon^2 h'/h}{\oneoverrho} =0,
  \label{GSE-general}
\end{equation}
where we have rescaled Bernoulli's function as
\begin{equation}
H \equiv -\epsilon^2\ln (\epsilon^2h),
\end{equation}
so that $h$ remains an order unity quantity in our calculation. Here
$\A$ is the Alfv\'en discriminant defined by
\begin{equation}
  \A \equiv \frac{M_A^{-2} -1}{\varpi^2 \rho},
  \label{calAdef}
\end{equation}
where
\begin{equation}
  M_A^2 \equiv \frac{\rho u^2}{B^2} = \oneover{\beta^2 \rho}
\end{equation}
is the Alfv\'en Mach number. Hence $\A$ is positive when the total
velocity is less than the Alfv\'en speed, and negative when the
total velocity is larger than the Alfv\'en speed.  From the form of
the GSE \eqref{GSE-general}, we see that the conserved angular
momentum flux $J$ is not freely specifiable. It is determined by the
condition of smooth Alfv\'en transition. In order for the solution
to remain continuous and differentiable, one must impose
\begin{equation}
    J = \varpi^2 \; \text{ whenever } \; \A = 0.\label{J-condition}
\end{equation}
The elimination of $\rho$ in the equations in favor of $\A$ is based
on numerical considerations, since $\rho$ will in general vary by
many orders of magnitude, while $\A$ only varies moderately. As we
argued in the previous section, the sound speed $\epsilon$ is likely
to be small.  In terms of these variables, the BE takes the
form
\begin{equation}
  \abs{\Del \psi}^2 + \oneover{\A^2}\prn{\frac{J}{\varpi^2} -1 }^2 +
  \frac{2\varpi^2\brc{\Veff +
  \epsilon^2 \ln \brk{\epsilon^2 h/(\oneoverrho)}}}{(\oneoverrho)^2}  =
  0.\label{BE-general}
\end{equation}

\subsection{The Cold Limit}
With $\A$ implicitly defined in the BE \eqref{BE-general}, the GSE
is a PDE of mixed type, which demands different numerical methods in
different regions (see Heinemann \& Olbert 1978 and Appendix
\ref{Character}). There are three relevant signal speeds (which we
term sonic, slow, and fast in the appendix) involved in an MHD flow
(see Jackson 1975 or Shu 1992).  The loci where the poloidal fluid
speed equals those signal speeds separate the flow into four
regions. As the poloidal velocity exceeds the sonic speed, the
governing GSE changes from elliptic to hyperbolic. A wise strategy
might start with the search of appropriate boundary conditions in
the disk where $u_p^2 = 0$, and at the sonic surface (whose location
is still undetermined), followed by a standard scheme (e.g.,
relaxation) to obtain the interior solution. Beyond the sonic
surface, the GSE becomes hyperbolic. The boundary condition on the
sonic surface now serves as the initial condition, which we use to
integrate forward along characteristics toward the slow surface.  We
then follow similar procedures to obtain solutions from the slow
surface to the fast surface, and beyond.

A significant simplification can be achieved when the sound speed is
negligible, as in the outer problem of the X-wind (see
\S 4 of Shu et al. 1994b). The governing equations are treated as
power series expansion in $\epsilon$.  The leading term in the GSE
\eqref{GSE-general} and the BE \eqref{BE-general} are
\begin{mathletters}
\begin{eqnarray}
  &&\Del \cdot (\A \Del \psi) + \oneover{\A}\prn{\frac{J}{\varpi^2}
  -1}\frac{J'}{\varpi^2} + \frac{{\beta^2}'\Veff}{(\beta^2 - \varpi^2 \A)^2}
  =0, \label{GSE}\\
  &&\abs{\Del \psi}^2 + \oneover{\A^2}\prn{\frac{J}{\varpi^2} -1 }^2 +
  \frac{2\varpi^2 \Veff}{(\beta^2 - \varpi^2 \A)^2} =
  0.\label{BE}
\end{eqnarray}
\end{mathletters}
Notice the lowest order term in $H$ vanishes independent of the form
of $h$.  In this limit, both the sonic speed and the slow speed
reduce to zero, and the first elliptic and hyperbolic parts of the
flow shrink down to the X-point. We are thus spared the vicissitudes
of this portion of the problem.  Once the fluid leaves the X-region
(with poloidal velocity greater than the slow speed), it proceeds to
the fast surface, where the governing equation becomes hyperbolic.

Najita \& Shu (1994) solved the GSE in the sub-Alfv\'enic region. By
introducing a generalized coordinate system, they were able to map
the location of the Alfv\'en surface to a known location, and
determine the functional form of $\beta(\psi)$ based on the position
and shape of the Alfv\'en surface. Their numerical scheme to find
$\beta (\psi)$ by iteration encountered a systematic ``drift
problem", however, and an artificial ``Alfv\'en seam'' was invented
to cope with this difficulty.

In a later treatment by Shang (1998), $\beta(\psi)$ was
specified in advance, limited in its functional form by
considerations of how the gas exits the X-region, an analysis that
we repeat in \S 4.2 (see also \S 5).  The GSE was not solved as a
PDE, but rather as an error estimator in a Weber-Davis type of
analysis, where $\psi$ as a trial function of spatial location is
obtained by interpolating between the known analytic forms in the
X-point neighborhood (see \S 4.1) and at asymptotic infinity (see \S
4.3).  The interpolation formula has a number of degrees of freedom,
which are adjusted to give ``least error'' in some sense when the
trial solution for $\psi$ is substituted back into the GSE.  The
rest of the problem, including the constraints of the conserved
quantities and smooth passage through the Alfv\'en and fast
surfaces, are performed exactly.  She verified the result derived by
Goldreich and Julian (1970) that passage through the Alfv\'en surface
is automatic in such a scheme if one has guaranteed it through the
fast surface. In fact, \S 5.2 demonstrates the
falsity of the frequent claim made otherwise in the literature that
$J(\psi)$ is set at the Alfv\'en surface; the claim holds only if one
already has a solution such that the wind passes smoothly through
the fast surface.


\section{Variational Principle}
Based on the above arguments, the X-wind is a fierce mathematical
beast, and a direct numerical attack is unlikely to subdue it fully.
To construct a global solution of the X-wind that accelerates
elements of plasma from the disk to super-magnetosonic speeds, we
must resort to a different approach.  Consider the following action
written down by inspection:\footnote{In their pioneering development
of magnetized stellar winds in a Grad-Shafranov formalism, Heinemann
\& Olbert (1978) noted in passing that the resultant equations could
be derived from a principle of least action, which differs in
detailed form from that used in this paper, but is in the same
spirit. However, they, and subsequent workers who have made similar
observations, did not exploit the principle to obtain actual wind
solutions.}
\begin{equation}
  S = \int \brc{\ahalf \mathcal{A} |\Del \psi|^2
  -\oneover{2\mathcal{A}} \prn{\frac{J}{\varpi^2} -1}^2 +
  \frac{V_{\text{eff}} + \epsilon^2 \ln [\epsilon^2 h/(\oneoverrho)]
  - \epsilon^2}{\beta^2 - \varpi^2 \mathcal{A}}} d^3x,
  \label{action}
\end{equation}
It is straight forward to demonstrate that variation against $\psi$
yields the GSE \eqref{GSE-general}, while variation against $\A$
gives the BE \eqref{BE-general}.  The challenges of constructing
solutions to a nonlinear PDE of mixed type is now transformed to
tuning trial functions of $\psi$ and $\mathcal{A}$ until a
local extremum of the action \eqref{action} is reached.

To formulate a scheme that is easy to implement numerically, we
consider a change of independent variables from the usual
cylindrical coordinates
\begin{displaymath}
  (\varpi, z, \varphi) \rightarrow (\varpi, \psi, \varphi).
\end{displaymath}
For a given value of $\psi$, the functional form of $z(\varpi)$
determines the shape of the given streamline, and
$\mathcal{A}(\varpi)$ offers information on the velocity distribution
along that streamline. Written in these new coordinates, and taking
the cold limit as $\epsilon \rightarrow 0$, the action reads
\begin{equation}
  S = 2\pi\iint \brc{\frac{1}{2} \A \prn{\pd{z}{\psi}}^{-1}
  \brk{1 + \prn{\pd{z}{\varpi}}^2}
  -\frac{1}{2\A}\pd{z}{\psi} \prn{\frac{J}{\varpi^2} -1}^2 +
  \pd{z}{\psi}\frac{V_{\text{eff}}}{\beta^2 - \varpi^2 \A}} \varpi
  d\psi d\varpi \label{new_action}
\end{equation}
Since $\A$ only enters the action as a constraint rather than a
dynamic variable (i.e., its derivative is absent in the action),
variation with respect $\A$ yields the BE as before, but now written
in a different set of coordinates,
\begin{equation}
  \prn{\pd{z}{\psi}}^{-2} \brk{1 + \prn{\pd{z}{\varpi}}^2}
  + \oneover{\A^2} \prn{\frac{J}{\varpi^2} -1}^2 +
  \frac{2\varpi^2\Veff}{(\beta^2 - \varpi^2 \A)^2} = 0
\end{equation}
Variation with respect to $z$ gives
\begin{eqnarray*}
  &&\delta S_z = 2\pi\iint \Bigg \{-\frac{1}{2} \A \prn{\pd{z}
  {\psi}}^{-2}
  \pd{\delta z}{\psi}
  \brk{1 + \prn{\pd{z}{\varpi}}^2}+\A \prn{\pd{z}{\psi}}^{-1}
  \prn{\pd{z}{\varpi}} \pd{\delta z}{\varpi}\\
  &&\qquad -\frac{1}{2\A}\pd{\delta z}{\psi} \prn{\frac{J}{\varpi^2}
  -1}^2 +
  \pd{\delta z}{\psi}\frac{V_{\text{eff}}}{\beta^2 - \varpi^2 \A}
  +\pd{z}{\psi}\frac{V_{\text{eff},z}}{\beta^2 - \varpi^2 \A}
  \delta z \Bigg \} \varpi
  d\psi d\varpi.
\end{eqnarray*}
Integrating by parts, we have
\begin{eqnarray*}
  &&\delta S_z = 2\pi\int \brc{-\frac{1}{2} \A \prn{\pd{z}{\psi}}^{-2}
  \brk{1 + \prn{\pd{z}{\varpi}}^2}
  -\frac{1}{2\A}\prn{\frac{J}{\varpi^2}
  -1}^2 + \frac{V_{\text{eff}}}{\beta^2 - \varpi^2 \A}}
  \varpi \delta z \Bigg \vert_{\psi = 0}^{\psi =1}d\varpi\\
  &&\qquad +2\pi \int\A \prn{\pd{z}{\psi}}^{-1}
  \prn{\pd{z}{\varpi}} \delta z \varpi \Bigg \vert_{\varpi =
  1}^{\varpi = \infty}
  d\psi\\
  &&\qquad +2\pi\iint \pd{}{\psi}\Bigg \{\frac{1}{2} \A
  \prn{\pd{z}{\psi}}^{-2}
  \brk{1 + \prn{\pd{z}{\varpi}}^2}+\frac{1}{2\A}\prn{\frac{J}
  {\varpi^2}
  -1}^2 - \frac{V_{\text{eff}}}{\beta^2 - \varpi^2 \A}
  \Bigg \} \varpi \delta z d\psi d \varpi\\
  &&\qquad +2\pi \iint \Bigg\{
  \pd{z}{\psi}\frac{\varpi V_{\text{eff},z}}{\beta^2 - \varpi^2 \A}
   - \pd{}{\varpi} \brk{\A \prn{\pd{z}{\psi}}^{-1}
  \prn{\pd{z}{\varpi}}\varpi }\Bigg \} \delta z
  d\psi d\varpi.
\end{eqnarray*}
Since we specify the boundary condition $z = 0$ on $\psi = 0$, and
$z = Z(\varpi)$ on $\psi = 1$, where $Z(\varpi)$ is a known
function, we see that $\delta z$ vanishes on these two boundaries.
As we shall see later \citep[see ][]{paper2, paper5}, the
solution near the X-point and asymptotically can be constructed
analytically.  Thus $\delta z$ also vanishes when $\varpi =1$ and
$\varpi \rightarrow \infty$ in our variational scheme, and both
surface terms vanish in the above expression.  In order for the
action to be stationary against any choice of $\delta z$, the
solution must satisfy the Euler Lagrange equation,
\begin{eqnarray}
  &&\frac{\delta S}{\delta z} =  \pd{}{\psi}\brc{\frac{1}{2} \A
  \prn{\pd{z}{\psi}}^{-2}
  \brk{1 + \prn{\pd{z}{\varpi}}^2}+\frac{1}{2\A}\prn{\frac{J}
  {\varpi^2}
  -1}^2 - \frac{V_{\text{eff}}}{\beta^2 - \varpi^2 \A}
  } \varpi \nonumber \\
  &&\qquad + \pd{z}{\psi}\frac{\varpi}{\beta^2 - \varpi^2 \A}
  \pd{V_{\text{eff}}}{z}
   - \pd{}{\varpi} \brk{\A \prn{\pd{z}{\psi}}^{-1}
  \prn{\pd{z}{\varpi}}\varpi } = 0.
\label{dSdz}
\end{eqnarray}
Dividing both sides by $\varpi$, we can simplify the Euler-Lagrange equation
\eqref{dSdz} to obtain
\begin{eqnarray}
  &&-\ahalf \pd{\A}{\psi}\brc{
  \prn{\pd{z}{\psi}}^{-2}
  \brk{1 + \prn{\pd{z}{\varpi}}^2}+ \frac{1}{\A^2}
  \prn{\frac{J}{\varpi^2}
  -1}^2 + \frac{2\varpi^2 \Veff}{(\beta^2 - \varpi^2 \A)^2}} \nonumber \\
  &&\qquad + \frac{1}{\A}\prn{\frac{J}{\varpi^2}
  -1} \frac{J'}{\varpi^2}  + \frac{{\beta^2}' \Veff}{(\beta^2
  - \varpi^2 \A)^2} - \oneover{\varpi}\prn{\pd{z}{\psi}}^{-1}
  \pd{}{\varpi} \brk{\A
  \prn{\pd{z}{\varpi}}\varpi } \nonumber\\
  &&+ \prn{\pd{z}{\psi}}^{-1}\pd{}{\psi}\brc{\A
  \prn{\pd{z}{\psi}}^{-1}
  \brk{1 + \prn{\pd{z}{\varpi}}^2}}= 0.\label{GSE_halfdo}
\end{eqnarray}
We notice that the coefficient of $\partial \A/\partial \psi$ is
simply the BE, which vanishes at a local extremum of the action. One
may easily check that the other terms yield the conventional GSE,
but written in our new coordinates.

\section{Boundary Conditions}
\subsection{X-point}\label{X-point}
With the new coordinates, the computational domain is bounded by
$\psi \in [0, 1]$ and $\varpi \in [1, \infty)$.  The X-point in
these coordinates is a singularity given by $\varpi = 1$ for all
values of $\psi$.  Fortunately, we have analytic solutions there.
From this point onward, we shall work with a scaled Alfv\'en
discriminant
\begin{equation}
  \chi \equiv \frac{\A}{\beta^2}.
\end{equation}
This function has the advantage of remaining finite even when
$\beta$ diverges. For a given functional form of $\beta$ (which
tells us how matter is loaded onto the field lines), the Alfv\'en
discriminant has the series expansion in $\varpi-1$.
\begin{equation}
  \chi_X = 1 + \chi_1(\psi) (\varpi-1) + \chi_2(\psi) (\varpi-1)^2 + ...,
\end{equation}
Here a subscript $X$ reminds us that this series solution is valid
near the X-point.  In order to match asymptotically onto the outer
limit of the inner problem (Shu et al. 1994b), the density $\rho \equiv
\beta^{-2}(1- \varpi^2 \chi)^{-1}$ must diverge as $(\varpi-1)^{-2}$
near the X-point. This requirement translates to $\chi_1 = -2$ for
all values of $\psi$. Similarly, for the coordinate $z$ (now a
dependent variable), we expand it as,
\begin{equation}
    z_X = z_1(\psi) (\varpi-1) + z_2(\psi)(\varpi-1)^2 + ....
\end{equation}
Notice that the Jacobian near the X-point is
\begin{displaymath}
  \sqrt{g} = z_1'(\varpi-1) \; \text{ as } \; \varpi \rightarrow 1.
\end{displaymath}
which is expected since the entire line of $\psi \in [0, 1]$ is
mapped into a single point, and the Jacobian must vanish in this
situation. Substituting the series expansion into the GSE
\eqref{GSE_halfdo} in the transformed coordinates, the lowest order
is $(\varpi-1)^{-2}$.
\begin{displaymath}
  \pd{}{\psi}\brk{\frac{\beta}{z_1'}
  \prn{1+z_1^2}} = 0,
\end{displaymath}
which has the solution
\begin{equation}
  z_1 = \tan \vartheta, \qquad \vartheta =
  \frac{1}{K} \int_0^\psi \beta d\psi.\label{-2nd_order}
\end{equation}
If we assume that the upper boundary of the X-wind ($\psi = 1$) near
the X-point forms an angle of $\theta_X$ with the $x$ axis, we have
\begin{displaymath}
  \tan \theta_X \approx \frac{z}{\varpi-1} \Big \vert_{\chi
  \rightarrow 1} = z_1(\psi = 1) = \tan \frac{1}{K} \int_0^1 \beta
  d\psi.
\end{displaymath}
Thus, the integration constant is given by
\begin{equation}
  K = \frac{\bar \beta}{\theta_X}, \qquad \bar\beta = \int_0^1
  \beta d\psi.
\end{equation}
Substituting back into the Bernoulli's equation allows us to solve
for $\chi_2$,
\begin{equation}
  \chi_2 = 3- \frac{\sqrt{4\cos^2 \vartheta -1}}{K \beta \cos^2
  \vartheta} \label{chi_2}
\end{equation}
For better numerical accuracy, we carry out the computation to next
order.  The next term in the series expansion of the GSE is $O[(\varpi-1)^{-1}]$:
\begin{displaymath}
  \pdd{\mathcal{Q}}{\vartheta}  + \mathcal{Q} = \sin
  \vartheta,
\end{displaymath}
where $\mathcal{Q} = z_2\cos^3\vartheta$. Given the boundary
condition $z=0$ at $\psi=0$ for all values of $\chi$, the above
equation may be solved to give
\begin{equation}
  z_2 = \ahalf \sec^2\vartheta \prn{q \tan \vartheta - \vartheta
  }.\label{z_2}
\end{equation}
The integration constant $q$ can be determined by expanding the
upper boundary near the X-point.  Substituting into the BE, we can determine the last
term without the knowledge of $J$ as
\begin{equation}
  \chi_3 = \frac{\cos^2\vartheta -2 + \tan \vartheta(q \tan
  \vartheta - \vartheta)}{2K \beta \cos^2 \vartheta \sqrt{4
  \cos^2 \vartheta -1}} + \frac{\sqrt{4 \cos^2 \vartheta -1}
  (4-q \sec^2\vartheta)}{2K \beta \cos^2 \vartheta} - 4. \label{varpi_3}
\end{equation}

\subsection{Specifying Mass Loading and Difficulties with the
Boundary Layer} \label{Boundary_Layer} Since the X-wind is driven
magnetocentrifugally, one would naively expect that it is bounded
away from the polar axis (at least in the immediate vicinity of the
X-point) by some curve which intersects the disk.  In the outer
limit of the inner problem (see eq. 3.10d of Shu et al. 1994b), the gas
pressure $p = \epsilon^2 \rho$ takes the form
\begin{displaymath}
  p \rightarrow \brk{\frac{\bar \beta}{\vartheta_x(0) \beta}}
  \sigma^{-2} (4\cos^2\vartheta -1)^{-1/2}, \; \text{ as } \; \sigma
  \rightarrow \infty,
\end{displaymath}
where $\tan\vartheta = z/(\varpi-1)$.  As $\vartheta_x \rightarrow
\pi/3$ (which is the critical angle for the last matter carrying
streamline), the pressure diverges unless
\begin{displaymath}
  \beta \propto (4 \cos^2\vartheta -1)^{-1/2}, \; \text{ as } \; \psi
  \rightarrow 1.
\end{displaymath}
Substitute this functional form into the lowest order equation
\eqref{-2nd_order}, finite magnetic field and pressure on the last
streamline demands
\begin{equation}
  \beta \propto (1-\psi)^{-1/3}, \;\text{ as }\; \psi \rightarrow 1.
\end{equation}

The divergence of $\beta$ should not come as a surprise.  Recall
that the last streamline is defined to be the boundary between the
X-wind and the dead zone.  To ensure analyticity across this
boundary, we must have $\rho \rightarrow 0$ as $\psi \rightarrow 1$.
Now since neither the magnetic field nor the velocity become
singular, we must take $\beta \rightarrow \infty$ on that last
streamline, so that the product $\beta^2 \rho^2 = B^2/u^2$ remains
finite.  With this limit in place, we see that the rescaled Alfv\'en
discriminant
\begin{displaymath}
  \chi = \frac{1 - 1/\beta^2 \rho}{\varpi^2} \rightarrow
  \oneover{\varpi^2}
\end{displaymath}
remains positive for all points along the last streamline.  In other
words, the flow on the last streamline is always sub-Alfv\'enic,
since the Alfv\'en speed is infinite there.  This behavior of the
last streamline requires a double limiting procedure if we were to
accurately construct the asymptotic solution on that interface.  We
speculate that this difficulty is an indication that the last
streamline needs to be treated as a boundary layer.  This
speculation is reinforced by the fact that the last X-wind
streamline is the outer bounding surface to a sheet of axisymmetric
current defined by opened stellar field lines that reverse poloidal
directions as the current sheet is crossed and we find ourselves in
the dead zone of the overall X-wind/funnel-flow configuration (see
Fig. 1).  Until we actually construct such a
boundary-layer/current-sheet theory, we adopt a simple modification
to deal with the problem:  we truncate the formal wind solution at
some $\psi_1<1$, below which $J$ and $\beta$ remain finite.  We then
add the part between $\psi_1$ and $1$ to the dead zone fields of the
problem, i.e., treat the last few streamlines as opened vacuum
fields and impose the pressure balance condition at $\psi_1$.

\subsection{Asymptotic Solution}\label{Asymptotic Solution}
The asymptotic solution at large distances from the X-point was
constructed by Shu et al. (1995).  In particular, for a wind
reaching more or less constant terminal velocity, its density scales
roughly as $\rho \propto \varpi^{-2}$, and the Alfv\'en discriminant
$\chi \rightarrow -1/\beta^2 \rho \varpi^2$ is a slowly varying
function of $r$.  By ignoring all radial derivatives compared to
angular derivatives, the GSE and the BE admit solutions of the form
\begin{equation}
  \chi = - 1/\beta C, \qquad \sin \theta = \sech [C^{-1}I(C, \psi)],
  \qquad I(C, \psi) = \int_0^\psi \frac{\beta
  d\psi}{\sqrt{2J - 3- 2C \beta}},\label{sol_asymp}
\end{equation}
where $\theta$ is the usual polar angle in spherical coordinates,
and $C$ is a ``constant of integration'' that vary slowly in $r$. In
an inertial frame, the wind reaches a terminal velocity given by
\begin{equation}
  v_w = (2J-3-2C \beta)^{1/2}.
\end{equation}
To determine the constant $C$, we impose pressure balance between
the X-wind and the dead zone. Since the dead zone field lines carry
no inertia, they do not develop a toroidal component, and the
poloidal field satisfies the vacuum equation (Ostriker \& Shu1994).
Asymptotically, we do not expect the field lines to pinch toward the
rotational axis since the hoop stress is vanishingly small (Shu et
al. 1995).  For simplicity, we assume the boundary layer deviates
only slightly from a cylindrical surface at the asymptotic infinity
(an assumption which shall be checked a posteriori for consistency).
For any given (large) value of $r$, we can approximate the boundary
locally by $\varpi = $ const. Then a particular solution is
$\mathbf{B} = B_{\rm hc} \zhat$. For the hollow-cone region to trap
the same amount of {\it net} flux as the wind part and have a
cross-sectional area of $\pi \varpi_{\rm hc}^2$, we have
\begin{displaymath}
  B_{\rm hc} = \frac{2\phi_{\rm hc} \bar \beta}{\varpi_{\rm hc}^2},
\end{displaymath}
where $\phi_{\rm hc}$ is a number ranging from $1$ to $3$ depending
on the fraction of closed field lines in the dead zone (compare Fig.
1 of this paper with Fig. 1 of Shu et al. 2001).  The maximal case
$\phi_{\rm hc} = 3$ has three times as many field lines as the
minimal case $\phi_{\rm hc} = 1$, but the extra field lines cancel
in oppositely directed pairs and contribute no net flux.  The
overall solution does not depend sensitively on the number
$\phi_{\rm hc}$, and we take $\phi_{\rm hc}$ henceforth to be unity
as illustrated in Figure 1 of the current paper.

In contrast, the wind region is dominated by the toroidal field
\begin{displaymath}
  B_{w, \varphi} = \beta \rho u_\varphi = -\frac{J-\varpi^2}
  {\varpi^3 \chi \beta} \rightarrow -\frac{C}
  {\varpi}.
\end{displaymath}
The poloidal field $B_{w,p} = \beta \rho v_w \propto 1/\varpi^2$ is
much weaker in this limit.  By equating the magnetic pressures on
both sides of the boundary, $B_{\rm hc}^2 = B_{w, \varphi}^2$, we obtain
\begin{equation}
  C = \frac{2\bar \beta}{\varpi_{\rm hc}} = \frac{2\bar \beta}{r}
  \cosh [C^{-1} I(C, 1)],
\end{equation}
which implicitly defines $C(r)$.  Since $I$ only depends very weakly
on $C$, this expression shows that $C \rightarrow 0$ logarithmically
as $r \rightarrow \infty$.  Notice that this limiting behavior of
$C$ ensures that $\varpi_{\text{hc}}$ deviates from a constant only
logarithmically slowly, which validates our assumption on the
geometry of the boundary layer.  Written in our coordinates, the
asymptotic geometry of each streamline is given by
\begin{equation}
  z = \varpi \sinh[C^{-1} I(C, \psi)].
\end{equation}
In other words, each streamline is approximately radial, with a
logarithmic collimation toward the axis.

With $\rho \rightarrow C/\beta \varpi^2$ and $v_w \rightarrow $
constant, the poloidal and toroidal Alfv\'en speeds are given by
\begin{equation}
  v_{A,p}^2 = B_p^2/\rho \rightarrow C \beta v_w^2/\varpi^2, \qquad
  v_{A,\varphi}^2 = B_{\varphi}^2/\rho \rightarrow C\beta.
\end{equation}
Thus the Alfv\'en speed is dominated by the toroidal component,
which decreases logarithmically.  This means that the (poloidal)
terminal velocity is super-fast in the asymptotic regime, and the
wind has to make a fast mode transition along each streamline.  The
above analysis simply reiterates the claims made in Appendix
\ref{Character} that the asymptotic behavior of the flow is governed
by a hyperbolic differential equation.

\section{Global Solutions}
As a particular example, let us suppose that diffusive mass loading
onto field lines in the X-region produces a $\beta$ function which
has the form
\begin{equation}
  \beta = \frac{2}{3} \bar \beta (1-\psi)^{-1/3}.\label{beta}
\end{equation}
It is easy to verify that $\int_0^1 \beta d\psi = \bar \beta$. To be
definite, let us also assume that the upper boundary of the
X-wind near the X-point forms the maximum angle $\vartheta_X =
\pi/3$ with the equatorial plane in order for magnetocentrifugal
acceleration to operate. The $O[(\varpi-1)^{-2}]$ solution
\eqref{-2nd_order} takes the form
\begin{equation}
  z_1 = \tan \vartheta = \tan \frac{\pi}{3} \brk{1-(1-\psi)^{2/3}}.
  \label{-2nd_order_solution}
\end{equation}
In fact we have chosen a very special value for the opening angle.
Recall that the formal boundary between the X-wind and the dead zone
is characterized by vanishing $\rho$ with finite magnetic field.
That results in $\beta \rightarrow \infty$ and $\chi = \varpi^{-2}$.
If $\vartheta_X$ were smaller than $\pi/3$, one may check that the
boundary condition $\chi = \varpi^{-2}$ agrees with the series
solution of \S \ref{X-point} to the second order for all values of
$q$. This integration constant is computed by expanding the shape of
the last streamline near the X-point. However, when $\vartheta_X =
\pi/3$, the series solution agrees with the boundary condition only
if the quantity $q$ in equation (\ref{z_2}) satisfies
\begin{displaymath}
  q = \oneover{3} \prn{\frac{7}{4} + \frac{\pi}{\sqrt{3}}}.
\end{displaymath}
If $\vartheta_X > \pi/3$, the solution becomes discontinuous.  This
behavior is consistent with our physical intuition.  When the flow
is cold, the upper boundary of the X-wind is imposed by pressure
balance.  As one eases up the external pressure, $\vartheta_X$
increases. However, even when the external pressure drops to zero,
the matter carrying streamlines are confined to $\vartheta \le
\pi/3$, since it is the boundary where centrifugal effects can
overcome gravity. At least near the X-point, there is no freedom to
choose the shape of the last streamline.  Any excursion across this
boundary requires additional pressure support from the X-wind, which
calls for a warm rather than cold outflow.

In the particular example we are studying here, the second order
coefficient for $z$ becomes
\begin{equation}
  z_2 = \ahalf \sec^2 \vartheta \brk{\prn{\frac{7}{12} +
  \frac{\pi}{3\sqrt{3}}}
  \tan \vartheta - \vartheta}.
\end{equation}
For numerical tractability, we place the boundary layer at $\psi =
0.99$. Given the choice of mass loading in equation \eqref{beta},
the $\beta$ function is not much larger than unity there.

\subsection{Fixing the Free Function $J(\psi)$}
\label{find_J}
\begin{figure}[ht]
\begin{center}
\includegraphics[width = 4.5in, angle = 0]{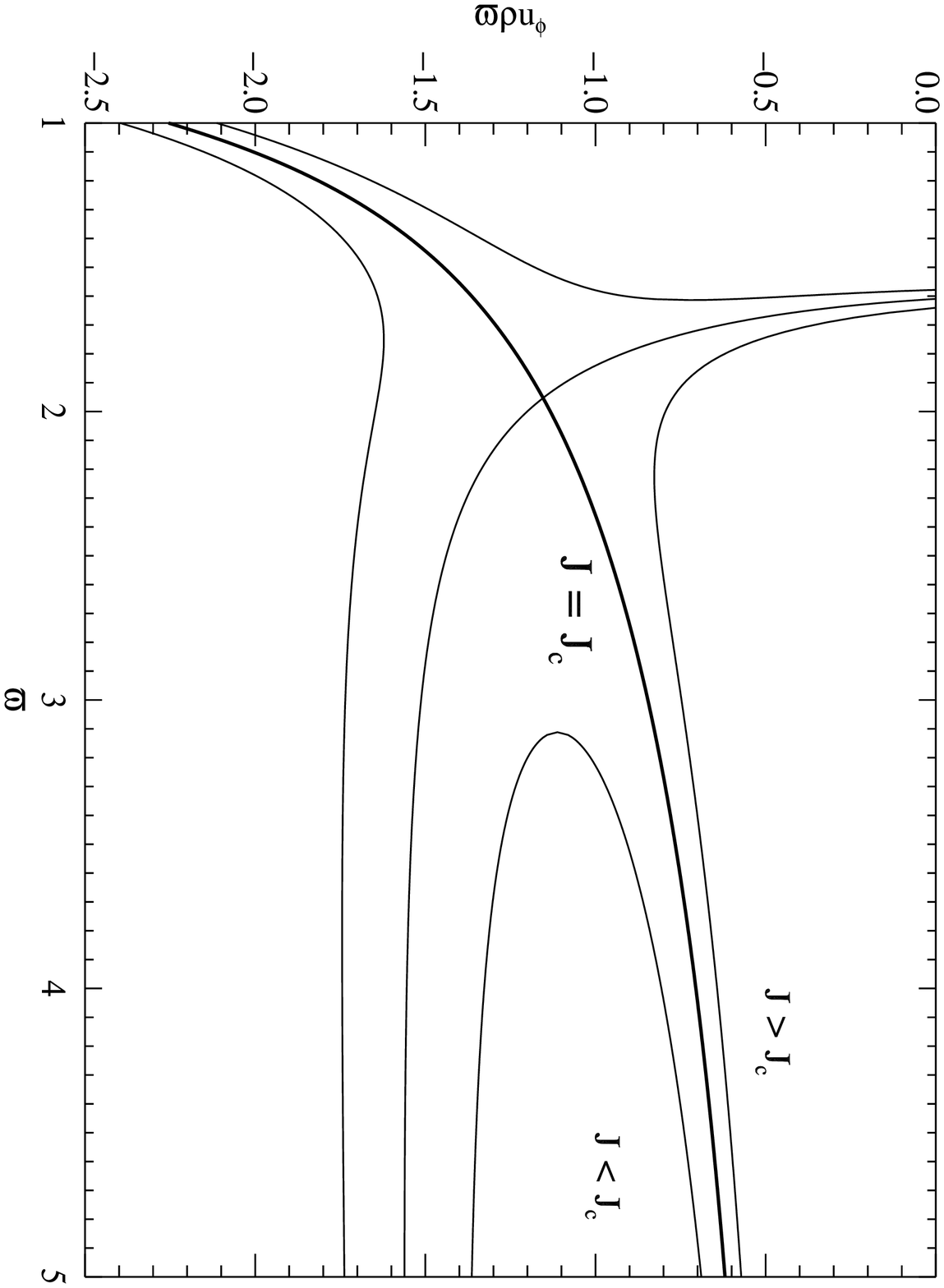}
\end{center}
\caption{Determination of the critical value of $J$ that allows a
fast mode transition}
\label{fast_crossing}
\end{figure}

The BE \eqref{BE} is actually
a quartic algebraic equation for the Alfv\'en discriminant once the
shape of the streamlines are known. The Alfv\'en surface here is not
a real singularity of the equation; it simply ensures that $\chi =
0$ is a solution when $J = \varpi^2$. In other words smooth crossing
of the Alfv\'en surface does not uniquely determine the value of
$J$.  To see this, let us define
\begin{displaymath}
  \mathcal{L} = \varpi \rho u_\varphi = - \oneover{\beta^2 \chi}
  \prn{\frac{J}{\varpi^2} -1}.
\end{displaymath}
It is always negative and asymptotes to zero for the wind since the
magnetic field lines form a trailing spiral.  The BE can be written
as
\begin{equation}
  (\abs{\Del \psi}^2 + \mathcal{L}^2)(\beta^2 \mathcal{L}
  + J - \varpi^2)^2 +
  2\varpi^2 \Veff \mathcal{L}^2 = 0.\label{mod_BE}
\end{equation}
As long as $J$ is larger than some critical value, there are always
real and finite solutions to this equation, which means the Alfv\'en
surface is automatically crossed.  On the other hand, the fast point
is a real critical point for the BE.  A smooth fast mode transition
demands the BE to have a double root at the critical point (see Fig
\ref{fast_crossing}). That means not only does the left hand side of
\eqref{mod_BE} need to vanish, its derivative with respect to
$\mathcal{L}$ must vanish as well.

After some algebra, these
requirements can be written as
\begin{equation}
  \mathcal{L} = \abs{\Del \psi}^{2/3}
  (J - \varpi^2)^{1/3} \beta^{-2/3}.\label{fast}
\end{equation}
This expression is simply a statement that at the fast point, the
poloidal fluid velocity is equal to the magnetosonic speed, which
for $\epsilon = 0$ is equal to the total Alfv\'en speed. Notice that
both equations \eqref{mod_BE} and \eqref{fast} are automatically
satisfied by $\mathcal{L} = (J-\varpi^2) = 0$. This solution,
however, is unphysical since it has a discontinuity on the Alfv\'en
surface when $\chi = 0$.  Substituting equation \eqref{fast} back in
to the BE \eqref{mod_BE}, we have
\begin{equation}
  \brk{\abs{\Del \psi}^{2/3}
  \beta^{4/3} + (J - \varpi^2)^{2/3}}^3 +
  2\varpi^2 \Veff= 0.\label{degen_root}
\end{equation}
For a given value of $J$, the solutions to this equation give the
locations where the BE has degenerate roots.  If $J<J_c$, equation
\eqref{degen_root} has no roots in the super-Alfv\'en region
($\varpi^2 > J$). If $J>J_c$, then equation \eqref{degen_root} has
two roots in the super-Alfv\'en part of the flow.  The desired
solution is obtained when $J=J_c$, and there is only one double root
occurring at the fast mode transition point (see Fig. 2).

\subsection{Interpolation Schemes and Numerical Strategy}
Our strategy is then to find interpolations between the X-point
solution in \S \ref{X-point} and the asymptotic solution of \S
\ref{Asymptotic Solution} so that the action \eqref{new_action} is
extremized. Since the action involves an integral extending to $r
\rightarrow \infty$, and the streamlines are approximately radial,
in general the action integral is infinite. In the X-wind problem,
however, the assumption of stationarity is an approximation that
must fail physically at very large distances from the X-point.  If
the flow extends all the way to spatial infinity, then steady state
cannot be established in finite time.  To make the practical aspect
of this problem manageable, we opt to truncate the action integral
at some finite spatial surface, and assume that the solution is
identical to the asymptotic solution beyond that point. Then the
interpolation requires the intermediate solution to join smoothly on
to the asymptotic solution at the boundary.  Since the parameter $C$
that appears in the asymptotic solution is purely a function of $r$,
it is natural to choose the boundary surface at $r = r_0 <\infty$.
Thus, along a given streamline labeled by $\psi$, the action
involves an integral over the range $\varpi \in [1, \varpi_\infty]$,
where
\begin{equation}
  \varpi_\infty = \frac{2 \bar \beta}{C} \frac{\cosh[C^{-1}I(C,
  1)]}{\cosh[C^{-1} I(C, \psi)]} \label{varpi_infty}
\end{equation}
Here $I$ is the integral defined in equation \eqref{sol_asymp}, and
the asymptotic value of $z$ is given by
\begin{equation}
  z_\infty(\varpi, \psi) = \varpi_\infty \sinh \brk{C^{-1} I(C, \psi)},
\end{equation}

Since the asymptotic behavior of the streamlines are predominantly
radial with a logarithmic collimation toward the pole, we may
approximate them by linear functions.  There is a large class of
basis functions in which $z(\psi, \varpi)$ can be expanded. To avoid
unphysical oscillations introduced by higher order polynomial
interpolations, we approximate $z$ by a cubic spline such that the
second derivative $z_{\varpi\varpi}$ is a continuous piecewise
linear function.
\begin{equation}
  z_{\varpi \varpi} = f_i + (\varpi-\varpi_i)\frac{f_{i+1} - f_i}
  {\varpi_{i+1} - \varpi_i} , \; \text{ for } \;
  \varpi_i \le \varpi < \varpi_{i+1},
\end{equation}
where $i = 0...N-1$, with $\varpi_0 = 1$ and $\varpi_N =
\varpi_\infty$. The boundary conditions on $z_{\varpi\varpi}$ read
\begin{equation}
  f_0 = 2z_2, \qquad f_N = 0.
\end{equation}
Direct integration yields \citep{Press:1992}
\begin{eqnarray}
  &&z = a y_i + b y_{i+1} + c f_i + d f_{i+1},\label{y-spline}\\
  &&z_\varpi = \frac{y_{i+1}-y_i}{\varpi_{i+1}-\varpi_i} -
  \frac{3a^2-1}{6}(\varpi_{i+1} -
  \varpi_i)f_i + \frac{3b^2 -1}{6} (\varpi_{i+1} - \varpi_i)
  f_{i+1},\label{yp-spline}
\end{eqnarray}
where
\begin{eqnarray*}
  &&a \equiv \frac{\varpi_{i+1} -\varpi}{\varpi_{i+1} - \varpi_i},
  \qquad b \equiv 1-a = \frac{\varpi-\varpi_i}{\varpi_{i+1}-\varpi_i},\\
  &&c \equiv \frac{1}{6}(a^3-a)(\varpi_{i+1}-\varpi_i)^2, \qquad d \equiv
  \frac{1}{6} (b^3-b)(\varpi_{i+1}-\varpi_i)^2,
\end{eqnarray*}
and $y_i \equiv z(\varpi_i)$ for that particular streamline.  The
$y_i$ are determined by demanding $z_\varpi$ is continuous
throughout the domain. Explicitly,
\begin{equation}
  \frac{\varpi_i-\varpi_{i-1}}{6} f_{i-1} + \frac{\varpi_{i+1} - \varpi_{i-1}}{3} f_i +
  \frac{\varpi_{i+1} - \varpi_i}{6} f_{i+1} = \frac{y_{i+1} - y_i}{\varpi_{i+1} -
  \varpi_i} - \frac{y_i - y_{i-1}}{\varpi_i -
  \varpi_{i-1}},\label{smooth_condition}
\end{equation}
which is a set of $N-2$ linear equations for the $N$ $y_i$.  The
boundary conditions $y_0 = 0$ and $y_N = z_\infty$ close the
equations, and allows unique determination of $y_i$ once $f_i$ are
given. Since we have information on the slope of the solution on
both boundaries, they impose two further constraints
\begin{eqnarray}
  &&z_1= \frac{y_{1}-y_0}{\varpi_{1}-\varpi_0} - \frac{1}{3}(\varpi_{1} -
  \varpi_0)f_0 - \frac{1}{6} (\varpi_{1} - \varpi_0)
  f_{1},\label{ypXconstraint}\\
  &&z_{\infty,\varpi}= \frac{y_{N}-y_{N-1}}{\varpi_{N}-\varpi_{N-1}}
  + \frac{1}{6}(\varpi_{N} -
  \varpi_{N-1})f_{N-1} + \frac{1}{3} (\varpi_{N} - \varpi_{N-1})
  f_{N}.\label{ypinftyconstraint}
\end{eqnarray}
To demonstrate the principles, we choose $N=3$, so that all the
$f_i$ are constrained.  For a given set of $\varpi_i$, the equations
\eqref{smooth_condition}, \eqref{ypXconstraint}, and
\eqref{ypinftyconstraint} form a set of four linear equations, which
can be solved by standard means.  We also define $\varpi_2$ by
\begin{equation}
  \frac{\varpi_2-\varpi_1}{\varpi_1 - 1} = \frac{\varpi_\infty -
  \varpi_2}{\varpi_2 - \varpi_1},\label{varpi_2}
\end{equation}
i.e., we demand that the interval between interpolation points to
increase exponentially.  Thus the shape of each streamline is
parameterized by a single variable, $\varpi_1$.

The action integral and the asymptotic solution can be treated as
solutions to a set of simultaneous ``ordinary'' differential
equations
\begin{equation}
  \od{S}{\psi} = \int_1^{\varpi_\infty(\psi)} L \varpi
  d\varpi,\qquad
  \od{I}{\psi} = \frac{\beta}{\sqrt{2J(\psi) - 3 - 2 C
  \beta(\psi)}},\label{num_int}
\end{equation}
subject to the boundary conditions
\begin{displaymath}
  S(0) = 0, \qquad I(0) = 0.
\end{displaymath}
Here $L$ represents the Lagrangian appearing in the action
\eqref{new_action}.  For each value of $\psi$, to compute the right
hand side of equation \eqref{num_int}, one needs the values of
$\varpi_1(\psi)$, $I(\psi)$ and $I(\psi_1)$, where $\psi_1 = 0.99$
is the label of the boundary layer discussed in \S
\ref{Boundary_Layer}.  Ideally, one would like to specify the shape
of the last streamline by fixing the values of $\varpi_1(\psi_1)$
and $I(\psi_1)$ as boundary conditions, and vary the function
$\varpi_1(\psi)$ in a constrained manner to achieve a local extremum
of the action. In practice, we find it more convenient to implement
a scheme where only $\varpi_1(\psi_1)$ is given, and $I(\psi_1)$ is
determined as an eigenvalue.  This approach allows more freedom in
the parameter search for the desired $\varpi_1(\psi)$.  With each
streamline fully parameterized, one can proceed to determine the
necessary value of $J(\psi)$ that allows a smooth fast mode
transition according to the procedure outlined in \S \ref{find_J}.

Once $J(\psi)$ and $I(\psi)$ are both known, we can easily solve the
BE \eqref{BE} as an algebraic equation along each streamline for
$\mathcal{L}$ using standard techniques such as Laguerre's method
(see Press et al. 1992).  In particular, note that we do {\it not}
actually use the extremal property of the action principle with
respect to $\cal A$ to attack the Bernoulli equation, but effect
direct solutions of it instead. Increased numerical accuracy
constitutes only one reason for a mixed procedure, where we do find
the extremal action through variations of $\psi$, or equivalently,
through variations of $z(\varpi, \psi)$, as a substitute for solving
the Grad-Shafranov equation. There is a yet more practical reason.
It turns out the the correct solution sits on a saddle, where the
extremal action is minimized by variations of $\psi$ but maximized
by variations of $\cal A$.  This combination makes a numerical
search for the extremal action extremely difficult to execute in
practice, perhaps even impossible, if the search is carried out in
the double-function space of allowable $\psi$ and $\cal A$.

One further obstacle to
overcome is that the action integral \eqref{new_action} is
logarithmically divergent at the X-point. Recall that the Jacobian
of the coordinate transformation vanishes at the X-point since it
maps the entire axis of $\varpi = 1$ onto a single point. A series
expansion of the Lagrangian using the series solution of section
\S\ref{X-point} shows that it diverges as
\begin{equation}
  L_* = \frac{K\beta}{2(\varpi-1)}.
\end{equation}
Fortunately, this term does not enter into the variation scheme, and
we may safely remove it as a counter term from the Lagrangian, as is
the standard practice in quantum-field theory.

Finally, the function $\varpi_1(\psi)$ is modeled by a Hyman
filtered spline \citep{Hyman:1983} interpolating over evenly spaced
control points $\psi_i \in [0, \psi_1]$.  The values of $\varpi_1$
at these control points, $\varpi_1(\psi_i)$, are the parameters we
can adjust in our variation scheme.  We restrict the parameter space
to that satisfies the condition that the streamlines do not cross
and that each streamline is monotonic. We then adopt a genetic
algorithm to search for a set of $\varpi_1(\psi_i)$ that gives a
local extremum of the action \eqref{new_action}.

\section{Numerical Results}
We compute the streamlines for three cases of average mass loading
corresponding to $\bar \beta = 1, 2, 3$.  In each case, we place the
outer boundary of the computational domain at a constant radius so
that it intersects the last streamline at $\varpi_\infty(\psi_1) =
20$ (which yields $C = 0.1 \bar \beta$).  After a multidimensional
search, we locate the desired set of control points that extremize
the action. They are tabulated in Table \ref{control_pts}, and the
function $\varpi_1(\psi)$ is interpolated between these points as
described in the previous section.
\begin{table}[ht]
\begin{center}
\begin{tabular}{|c|c|c|c|c|c|c|}
\hline $\bar \beta$ & $\varpi_1(0.0)$ & $\varpi_1(0.2)$
&$\varpi_1(0.4)$
&$\varpi_1(0.6)$ &$\varpi_1(0.8)$ &$\varpi_1(0.99)$\\
\hline
1.0 &29.687 & 29.817 & 23.281 & 18.809 & 8.310 &6.000 \\
2.0 &28.281 & 29.165 & 24.644 & 17.774 & 11.150 &6.000\\
3.0 &28.384 & 28.985 & 23.990 & 19.965 & 10.139 &6.000\\
\hline
\end{tabular}
\caption{Values of control points $\varpi_1(\psi_i)$ that yield a
local extremum of the action.  The last value $\varpi_1(0.99)$ is
fixed as a boundary condition.} \label{control_pts}
\end{center}
\end{table}

\begin{figure}[ht]
\begin{center}
\includegraphics[width = 2.9in, angle = 90]{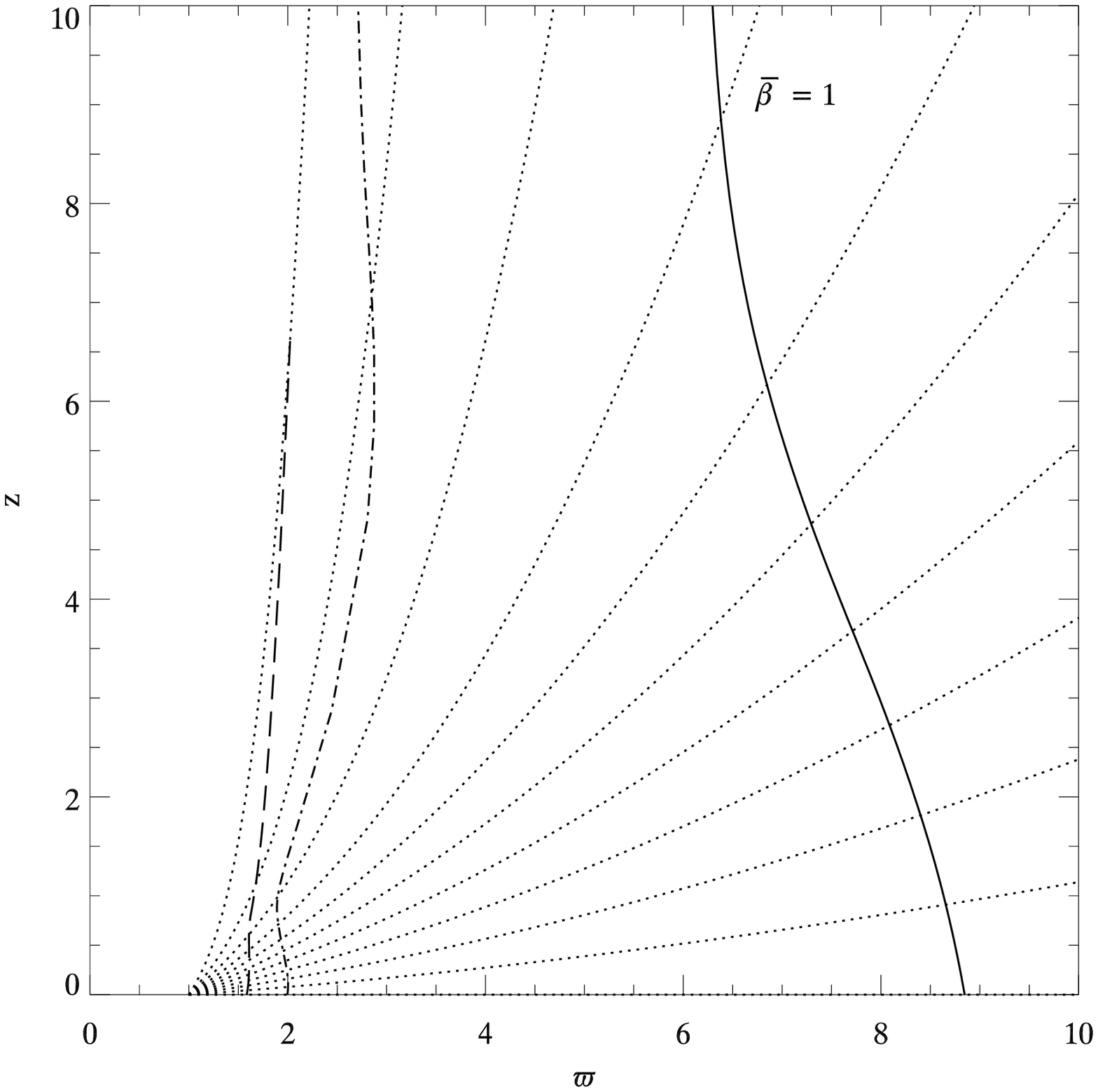}
\includegraphics[width = 2.9in, angle = 90]{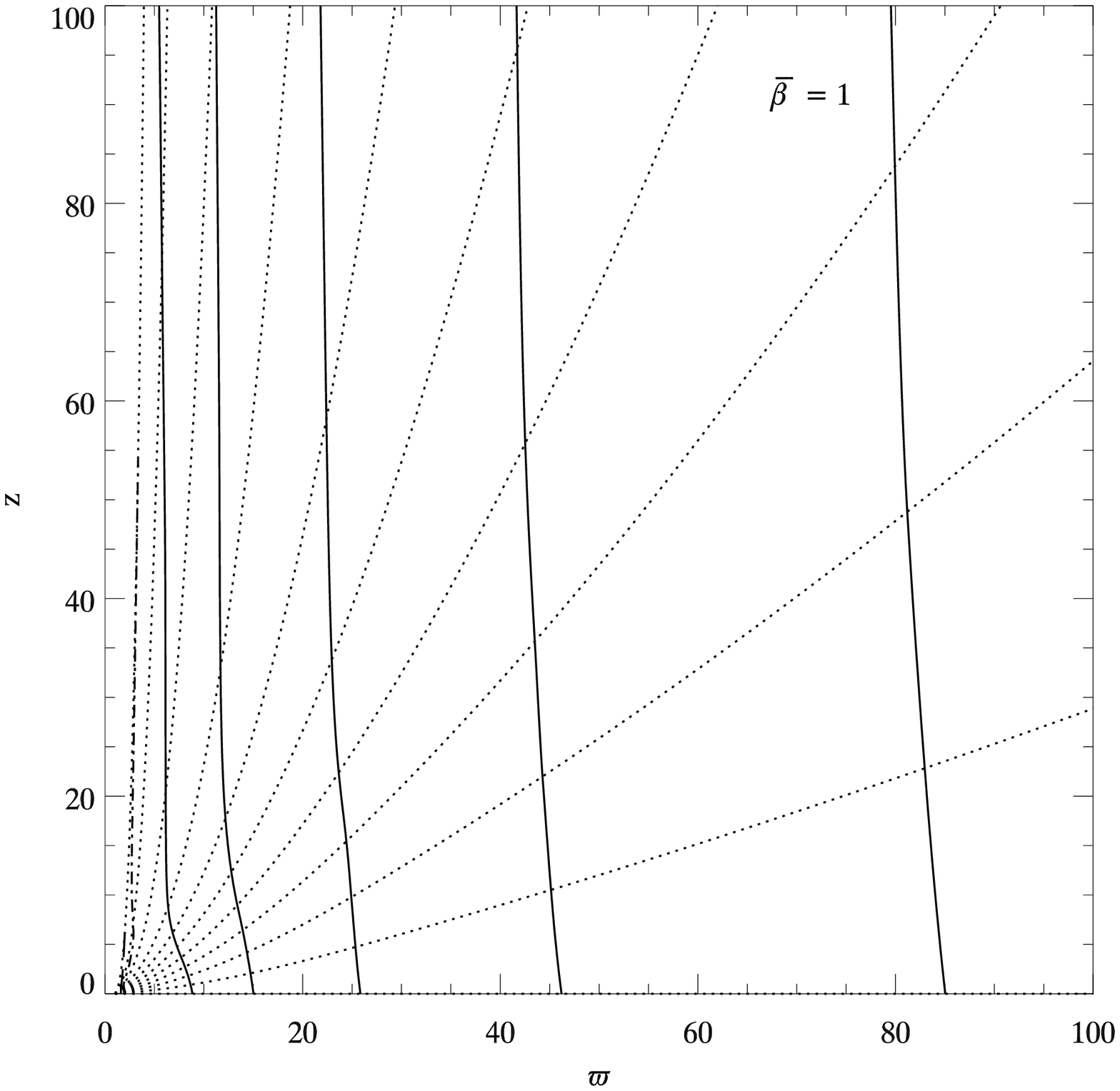}
\includegraphics[width = 2.9in, angle = 90]{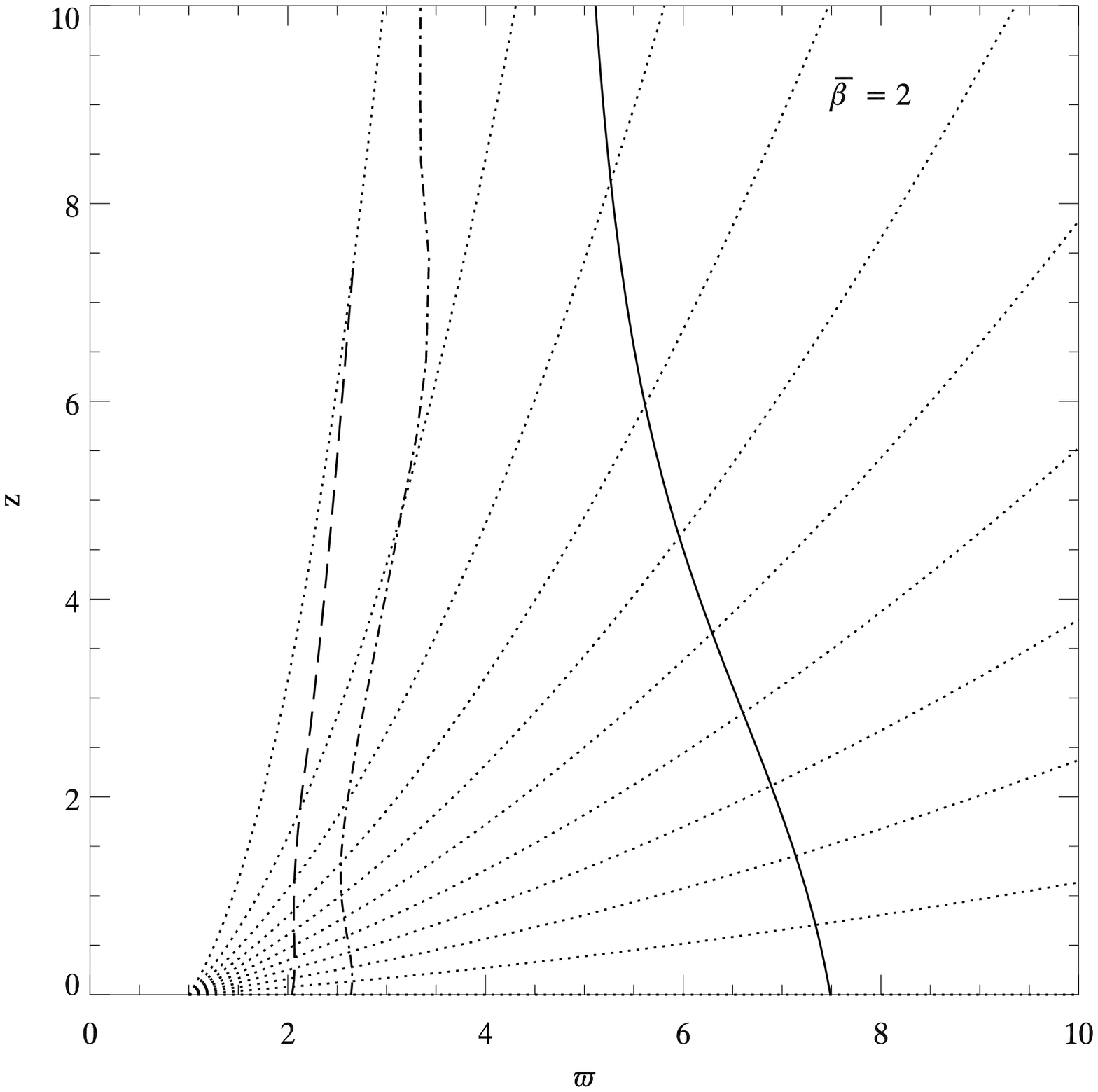}
\includegraphics[width = 2.9in, angle = 90]{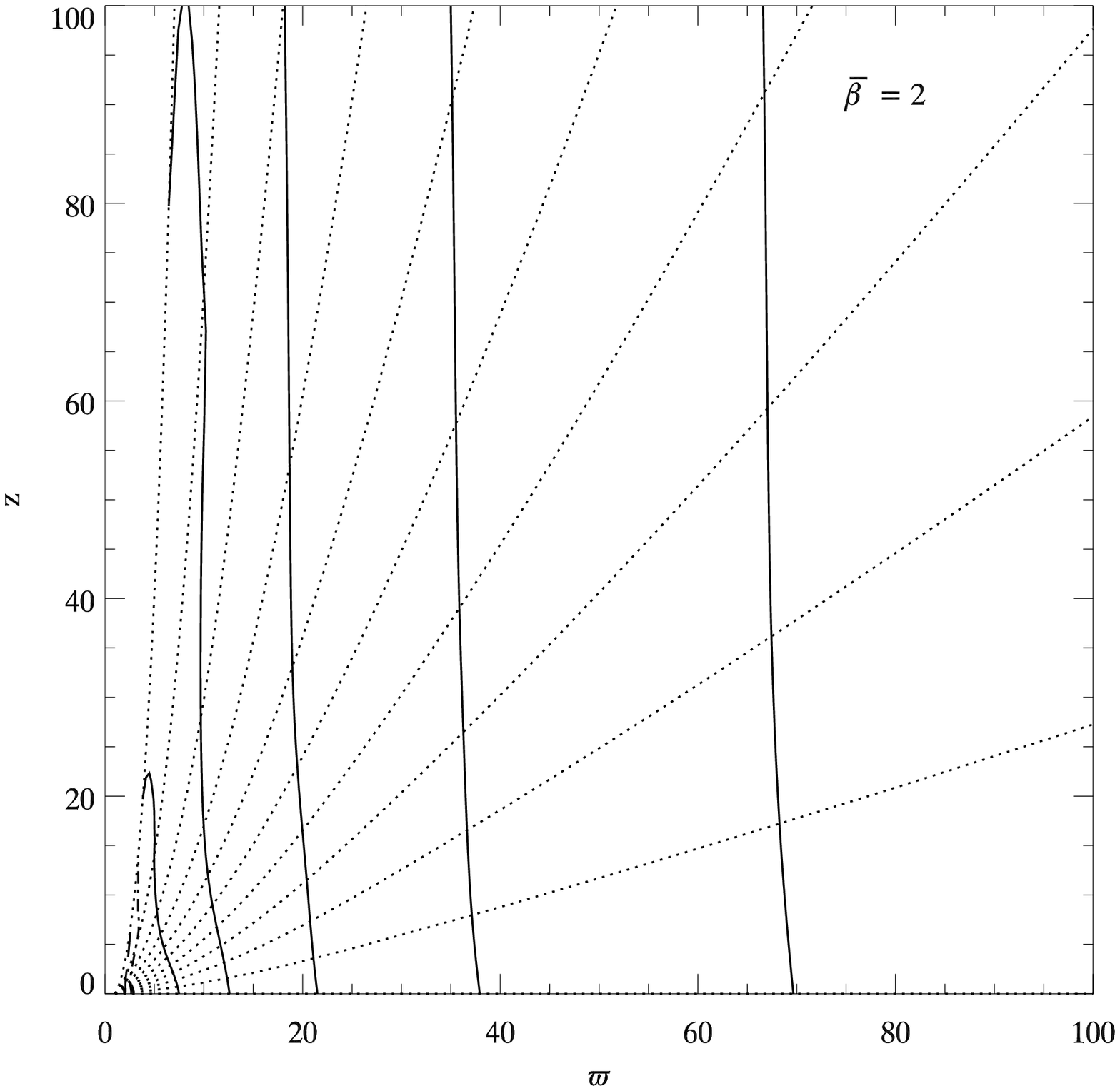}
\includegraphics[width = 2.9in, angle = 90]{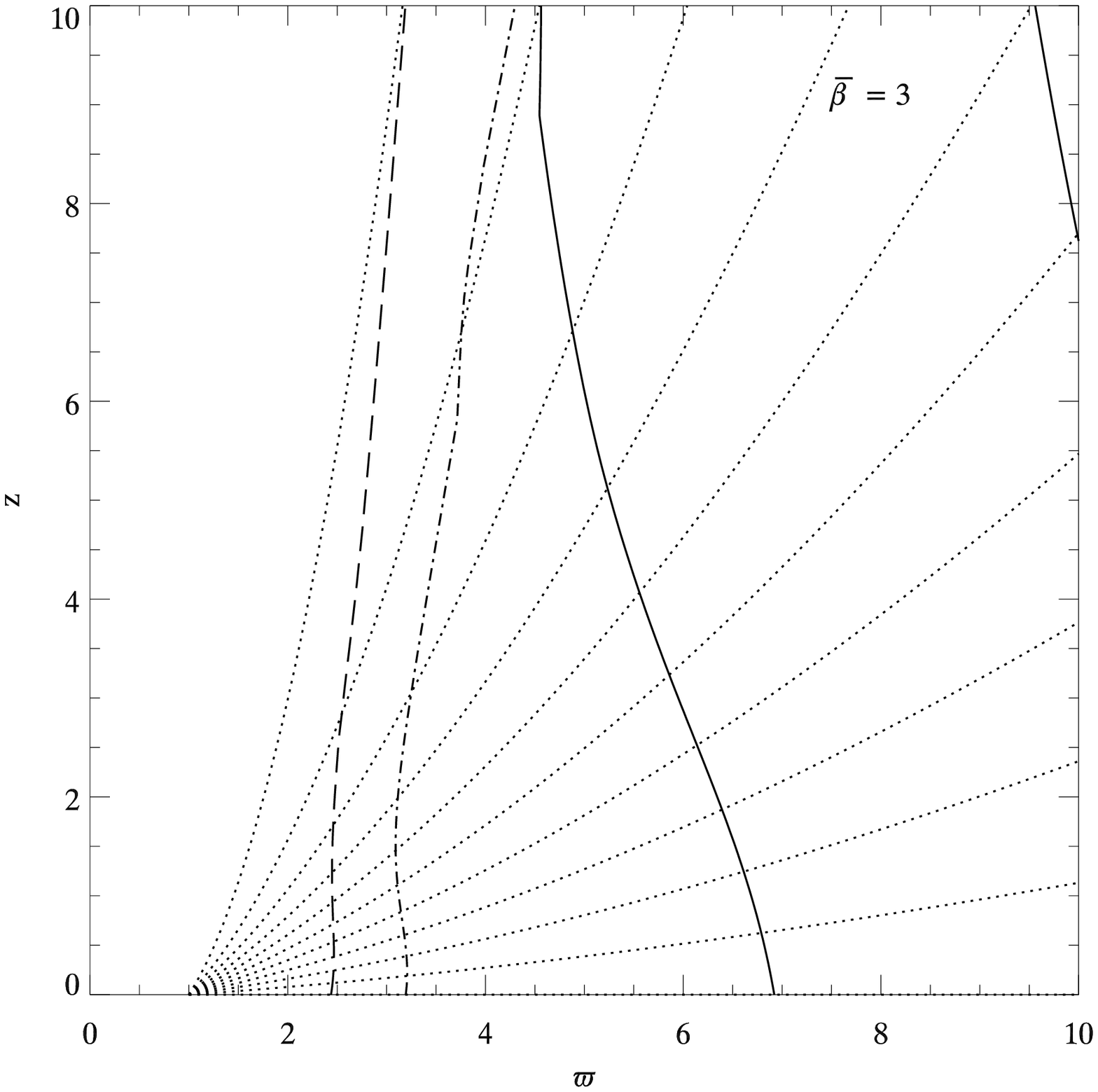}
\includegraphics[width = 2.9in, angle = 90]{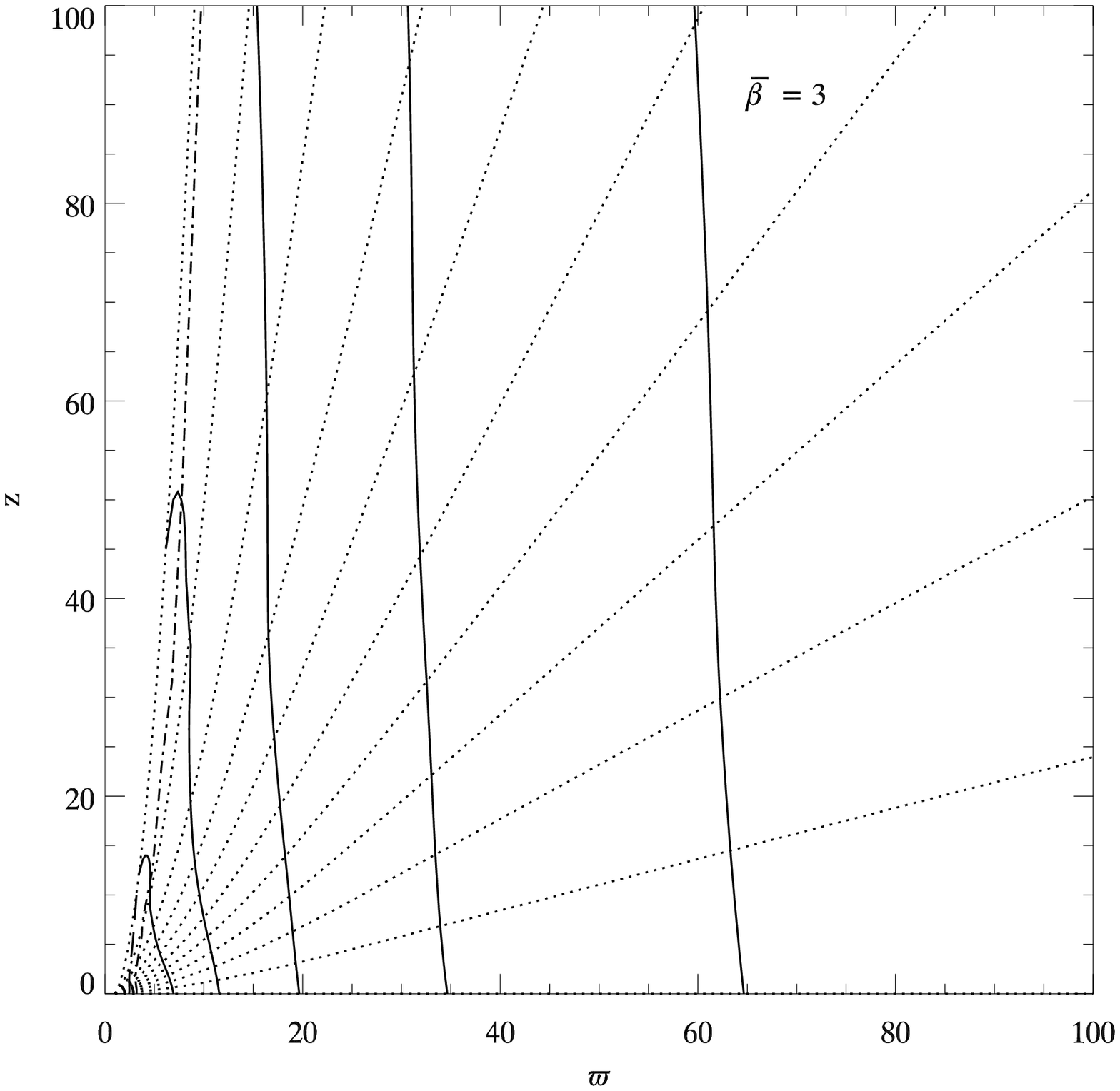}
\end{center}
\caption{Solutions for $\bar \beta = 1, 2, 3$.  The dotted curves
represent the streamlines labeled by constant $\psi$, and the solid curves
are isodensity contours separated by logarithmic intervals.  The dashed
curves are the location of the Alfv\'en surface in each case, and the
dashed-dotted curves mark the fast surface where the GSE becomes
hyperbolic.}
\label{crit}
\end{figure}

For each converged solution, we can numerically integrate the
asymptotic equation to evaluate $I(\psi)$.  For practical purposes,
we present here an interpolation formula that is a seventh degree
polynomial in $\beta^{-1}$, and the coefficients are tabulated in Table
\ref{I_inter}. Once $I(\psi)$ is known, one may determine the outer
boundary of the computational domain in accordance with the
asymptotic condition \eqref{sol_asymp}.  With the combination of
$\varpi_1(\psi)$ and $I(\psi)$, we are able to reconstruct the
streamlines with the spline interpolation scheme, and they are
depicted in Figure \ref{crit}. The location of the Alfv\'en surface
determines the value of $J$ as a function of $\psi$, which
ultimately allows us to compute the angular momentum being
transported as well as the terminal velocity along each streamline.
For convenience, we also present an interpolation formula for $J(\psi)$ as a
polynomial in $\beta$, with the coefficients tabulated in Table
\ref{J_inter}.
\begin{table}[ht]
\begin{center}
\begin{tabular}{|c|c|c|c|c|c|c|c|c|}
\hline
$\bar\beta$&$I_0$&$I_1$&$I_2$&$I_3$&$I_4$&$I_5$&$I_6$&$I_7$\\
\hline 1&0.732&-0.446&0.886&-0.511&-1.781&2.648&-1.397&0.263\\
2&0.842&0.413&-2.504&2.699&-11.768&24.849&-22.870&7.602\\
3&1.164&-2.915&30.674&-194.996&585.524&-998.495&$930.236$&$-347.764$ \\
\hline
\end{tabular}
\caption{Interpolation formula for $I_{\text{int}}(\psi) =
\sum_{i=0}^7 I_i \beta^{-i}$.  The interpolated function agrees with
the numerical values to within $0.5\%$.} \label{I_inter}
\end{center}
\end{table}

\begin{table}[ht]
\begin{center}
\begin{tabular}{|c|c|c|c|c|c|c|c|c||c|}
\hline
$\bar\beta$&$J_0$&$J_1$&$J_2$&$J_3$&$J_4$&$J_5$&$J_6$&$J_7$&$\bar J$\\
\hline
1&-2.791&17.214&-13.640&-10.294 & 22.483 &-13.656 &3.628 &-0.362 & 2.638\\
2&-14.944&45.162&-43.065&21.660&-6.265&1.057&-0.0969&0.00373 & 4.356\\
3&-20.285&40.676&-25.088&8.006&-1.440&0.148&-0.00825&0.000191& 6.202\\
\hline
\end{tabular}
\end{center}\caption{Interpolation formula for $J_{\text{int}}(\psi) =
\sum_{i=0}^7 J_i \beta^i$.  The last column gives the value $\bar J$
of $J(\psi)$ averaged over $\psi$ from 0 to 0.99.  The interpolated
function agrees with the numerical values to within $1\%$. }
\label{J_inter}
\end{table}

The solid lines in Figure 3 show the logarithmically spaced contours
of constant density.  It is evident that even though the dotted
streamlines become asymptotically radial and only collimate
logarithmically slowly, the density becomes cylindrically stratified
very quickly, giving the X-wind the illusion of a jet-like
appearance (Shang et al. 1998, 2002).

Detailed comparisons of the results obtained here with those given
by Shang (1998) show some differences, but the main impression is
how remarkably well the solutions obtained by the two very different
methods for the same mass-to-flux loading $\beta (\psi)$ agree with
one another.  Shang (1998) had a similar experience in comparing her
approximate, but analytic, solutions for the sub-Alfv\'enic region
to the exact, but numerical, solutions obtained by Najita \& Shu
(1994).

We attribute the fortunate circumstance to the following causes.  If
one is given somehow the geometric shape of the streamlines (or,
equivalently, the field lines in the meridional plane), then the
Weber-Davis procedure used by Shang, which includes an exact
solution of Bernoulli's equation, would give an exact solution of
the two-dimensional flow problem, provided one takes care to cross
each of the critical points properly.  In realistic circumstances,
the geometric shape of streamlines in the meridional plane is not
given a priori, but is to be found from the Grad-Shafranov equation
(or, equivalently, from minimizing the action by variations of the
stream function $\psi$).  However, if one has analytic solutions to
the Grad-Shafranov equation (from the work of Shu et al. 1994b and
1995) near and far from the X-point, then there are only so many
ways that one can adjust the function $z(\varpi, \psi)$ for values
of $\psi$ from 0 to 1 and of $\varpi$ close to 1 (or dimensionally,
$R_X$) to $\varpi \gg 1$ (or $R_X$) that will connect the shape of
the streamlines near the X-point (a fan) smoothly to those
appropriate at asymptotic infinity (radial outflow).  The procedures
used by Shang (1998) and those used here to make such adjustments
differ, but the global solution is relatively insensitive to these
details as long as one gets the conserved quantities: mass-to-flux
loading $\beta(\psi)$, angular momentum distribution $J(\psi)$, and
Bernoulli's constant $H(\psi) = 0$ correctly.

\section{Discussion and Conclusions}

\subsection{Recipe for Use of Results}

For the convenience of the reader, we summarize the recipes needed
to convert the results of the previous section into numerical X-wind
models for astronomical and meteoritical applications. Begin with
the equation that describes the dimensionless locus of a streamline
for given $\psi$ with numerical value between 0 and 1:
\begin{equation}
z = z(\varpi,\psi),
\end{equation}
where the functional form of $z(\varpi, \psi)$ is computed
numerically by the technique described in \S 5.2.

The reconstruction of streamline shapes, i.e., the function
$z(\varpi,\psi)$, is performed over three radial intervals whose end
points are $\varpi_0 \equiv 1$, $\varpi_1(\psi) > 1$,
$\varpi_2(\psi) > \varpi_1(\psi)$, and $\varpi_3(\psi) \equiv
\varpi_\infty(\psi) > \varpi_2(\psi)$ that give a geometrically
increasing separation:
\begin{equation}
  \frac{\varpi_2-\varpi_1}{\varpi_1 - 1} = \frac{\varpi_\infty -
  \varpi_2}{\varpi_2 - \varpi_1} ,
\label{separation-interval}
\end{equation}
where $\varpi_\infty (\psi)$ is given by equation
\eqref{varpi_infty}:
\begin{equation}
\varpi_\infty = \frac{2 \bar \beta}{C} \frac{\cosh[C^{-1}I(C,
1)]}{\cosh[C^{-1} I(C, \psi)]} .
\end{equation}
For practical computations, we choose $C = 0.1\bar \beta$ so that
$\varpi_\infty = 20$ on the $\psi =1$ streamline. The asymptotic
integral $I(C, \psi)$ in equation \eqref{sol_asymp} can be
approximated by a seventh degree polynomial in $\beta^{-1}$:
\begin{equation}
I(\psi) = I_0 + I_1 \beta^{-1}(\psi) + \dots + I_7 \beta^{-7}(\psi),
\end{equation}
where the coefficients $I_0$, $I_1$, $\dots$, $I_7$ are given in
Table 2 for the three values of $\bar \beta$ = 1, 2, 3. The function
$\varpi_1(\psi)$ represents the first nontrivial abscissa of the
spline beyond the X-point for each value of $\psi$ and is tabulated
in Table \ref{control_pts} for $\psi_i = 0.0, 0.2, 0.4, 0.6, 0.8,
0.99$.  For intermediate values, we interpolate $\varpi_1$ by a
piecewise cubic polynomial:
\begin{equation}
  \varpi_1 = h_0(\psi_i)+h_1(\psi)(\psi-\psi_i) +h_2(\psi_i)(\psi-\psi_i)^2
  +h_3(\psi_i)(\psi-\psi_i)^3 \text{ for } \psi_i
  \le \psi < \psi_{i+1}.
\end{equation}
In Table \ref{Hyman_coef}, we list the values of $h_j(\psi_i)$ for
each case of $\bar \beta$.  To get $\varpi_2(\psi)$ for any value of
$\psi$, one should use equation \eqref{separation-interval} after
first computing $\varpi_1(\psi)$ and $\varpi_\infty(\psi)$ at the
desired value of $\psi$.

\begin{table}[ht]
\begin{center}
\begin{tabular}{|c|c|c|c|c|c|c|}
\hline & $\psi$ & $0.0$ & $0.2$ & $0.4$ & $0.6$ &
$0.8$\\
\hline$\bar \beta = 1$& $h_0$&$29.687$&$29.817$ & $23.281$ & $18.809$ & $8.310$\\
& $h_1$&$17.315$&$-1.950$ & $-27.520$ & $-37.428$ & $-31.809$\\
& $h_2$&$-153.650$&$-333.100$ & $126.938$ &$-254.103$ &$103.428$ \\
& $h_3$&$351.625$ &$897.250$ &$-505.688$ &$893.830$ & $0.000$\\
\hline $\bar \beta = 2$ &$h_0$ & $28.281$ &$29.165$ & $24.644$ &
$17.774$ & $11.150$\\
&$h_1$ &$17.933$ & $-9.093$ & $-28.478$ & $-33.735$ & $-30.035$\\
&$h_2$ &$-67.563$ &$-105.763$ & $-61.800$ & $-9.272$ & $15.422$\\
&$h_3$ &$0.000$& $191.000$&$162.188$ & $61.737$ & $0.000$\\
\hline $\bar \beta = 1$&$h_0$ & $28.384$ & $28.985$ & $23.990$ &
$19.965$ & $10.139$\\
&$h_1$ & $16.995$ &$-9.015$ & $-22.550$ & $-34.628$ & $-35.107$\\
&$h_2$ & $-79.800$ & $-171.725$ & $96.763$ & $-215.142$ & $70.117$\\
&$h_3$ & $49.250$ &$459.625$ & $-423.188$ & $713.150$ & $0.000$\\
\hline
\end{tabular}
\caption{Interpolation coefficients for $\varpi_1(\psi)$}
\label{Hyman_coef}
\end{center}
\end{table}

The shape of each streamline given by $\psi$ = const in the three
radial intervals whose end points are $\varpi_0(\psi)=1$,
$\varpi_1(\psi)$, $\varpi_2(\psi)$, and
$\varpi_3(\psi)=\varpi_\infty(\psi)$ is then described by a
piecewise cubic polynomial, whose form, suppressing the implicit
dependence on $\psi$, is given by equation \eqref{y-spline}:
\begin{equation}
z(\varpi) = y_1a + y_2b+ {(\varpi_{i+1}-\varpi_i)^2\over 6}\left[
f_1(a^3-a)+f_2(b^3-b)\right],
\end{equation}
where
\begin{equation}
a\equiv {\varpi_{i+1}-\varpi\over \varpi_{i+1}-\varpi_i}, \qquad b
\equiv {\varpi-\varpi_i\over \varpi_{i+1}-\varpi_i}.
\end{equation}
The coefficients $y_1$, $y_2$, $f_1$, and $f_2$ are listed in Table
\ref{spline_coef} for discrete values of $\psi$ = 0.0, 0.2., 0.4,
0.6, 0.8, 0.99 in the three cases $\bar \beta = 1$, 2, 3. A Hyman
limited spline may be used to compute the streamlines for other
values of $\psi$.
\begin{table}[ht]
\begin{center}
\begin{tabular}{|c|c|c|c|c|c|c|c|}
\hline &$\psi$ &$0.0$ &$0.2$ &$0.4$ &$0.6$ &$0.8$ &
$0.99$\\
\hline $\bar \beta = 1$ &$y_1$ &$0.0$ & $12.764$ & $21.872$ & $41.255$ & $51.367$
&$525.487$\\
& $y_2$ & $0.0$ & $366.579$ & $627.01$ &$1044.15$ & $1402.6$ &$3648.96$\\
& $f_1$ & $0.0$ & $3.834 \times 10^{-3}$ &$3.083 \times 10^{-2}$ & $0.256$ & $4.120$
& $118.485$\\
& $f_2$ & $0.0$ & $-2.901 \times 10^{-5}$ & $-4.685 \times 10^{-4}$ & $-1.033\times 10^{-2}$
& $-0.301$ & $-48.051$\\
\hline $\bar \beta = 2$ &$y_1$ &$0.0$ &$12.091$ &$22.694$ &$29.659$ & $37.935$ & $73.230$\\
&$y_2$ &$0.0$ & $86.922$ & $156.801$ & $208.943$ & $267.766$ & $383.996$\\
& $f_1$ &$0.0$ & $2.359 \times 10^{-3}$ & $1.489\times 10^{-2}$ &
$9.705 \times 10^{-2}$ & $0.750$ & $9.943$\\
& $f_2$ &$0.0$ & $-1.991 \times 10^{-4}$ & $-1.518 \times 10^{-3}$ &
$ -1.333 \times 10^{-2}$& $-0.158$ & $-4.246$\\
\hline $\bar \beta = 3$ &$y_1$ & $0.0$ & $14.138$ & $22.415$ & $33.841$ & $26.286$ & $44.094$\\
&$y_2$ & $0.0$ & $82.541$ & $104.01$ & $133.299$ & $131.469$ & $173.659$\\
&$f_1$ & $0.0 $ & $1.903 \times 10^{-2}$ & $2.353 \times 10^{-2}$ & $5.056 \times 10^{-2}$
& $0.376$ & $2.950$\\
&$f_2$ & $0.0$ & $-3.651\times 10^{-3}$ & $-5.610 \times 10^{-3}$ &
$-1.722 \times 10^{-2}$ & $-0.102$ & $-1.424$\\
\hline
\end{tabular}
\caption{Spline coefficients for the
streamlines.}\label{spline_coef}
\end{center}
\end{table}

The partial derivatives of $\psi$ with $\varpi$ or $z$ are now given
by the usual rules of multivariate calculus:
\begin{equation}
\left( {\partial \psi \over \partial \varpi}\right)_z = -{(\partial
z/\partial \varpi)_\psi\over (\partial z/ \partial \psi)_\varpi};
\qquad \left( {\partial \psi\over \partial z}\right)_\varpi =
{1\over (\partial z/\partial \psi)_\varpi}.
\end{equation}
Table 3 gives $J(\psi)$ as a seventh order polynomial in $\beta(\psi)$:
\begin{equation}
J(\psi) = J_0 + J_1\beta(\psi) + \dots +J_7\beta^7(\psi),
\end{equation}
where $\beta(\psi)$ is itself given by
\begin{equation}
\beta(\psi) = {2\over 3}\bar \beta (1-\psi)^{-1/3},
\end{equation}
with $\bar \beta$ = 1, 2, 3 in the three chosen model cases. The
coefficients tabulated in Table 3 give a $J(\psi)$ that guarantees
that equation \eqref{BE},
\begin{equation}
\abs{\Del \psi}^2 + \oneover{\A^2}\prn{\frac{J}{\varpi^2} -1 }^2 +
  \frac{2\varpi^2 \Veff}{(\beta^2 - \varpi^2 \A)^2} =
  0,
\label{Bernoulli}
\end{equation}
has one real root for $\cal A$ in the computational domain when
$V_{\rm eff}$ is given by equation \eqref{Veff}:
\begin{equation}
\Veff = -\frac{1}{\sqrt{\varpi^2 + z^2}} -\ahalf \varpi^2 +
  \frac{3}{2}.
\end{equation}
By solving equation (\ref{Bernoulli}) as a fourth-order polynomial,
we may obtain the relevant value for the Alfv\'en discriminant $\cal
A$. Then the density can be computed through equation
\eqref{calAdef}
\begin{equation}
\rho = (\beta^2 - \varpi^2
\A)^{-1}.
\label{density}
\end{equation}
Note that this equation produces $\rho = \beta^{-2}$ at the
Alfv\'enic transition ${\cal A} = 0$.

With the density in place, we may obtain the two components of dimensionless
poloidal velocity from the definition \eqref{psidef} of $\psi$:
\begin{equation}
 u_\varpi \equiv \oneover{\varpi\rho} \pd{\psi}{z}, \quad
  u_z \equiv - \oneover{\varpi\rho} \pd{\psi}{\varpi}.
\end{equation}
The toroidal velocity in the corotating frame is given by equation \eqref{Jdef}
\begin{equation}
u_\varphi = {J(\psi)-\varpi^2\over \varpi (1-\beta^2\rho)}.
\label{toroidalvel}
\end{equation}
Note that $J(\psi) = \varpi^2$ where $\beta^2\rho = 1$ keeps
the toroidal velocity $u_\varphi$ well-behaved across the Alfv\'en surface, which is
not one of the critical surfaces of the overall problem.

The vector magnetic field may now be obtained from equation \eqref{betadef}:
\begin{equation}
{\bf B} = \beta \rho {\bf u},
\label{vectorB}
\end{equation}
whereas
the azimuthal velocity in the inertial frame is given by
\begin{equation}
v_\varphi = u_\varphi + \varpi ,
\label{inertialvphi}
\end{equation}
with the term $\varpi$ from the frame rotation being cancelled at
large $\varpi$ where $u_\varphi \rightarrow -\varpi$ because $\rho$
vanishes as $1/\varpi^2$ at large distances from the rotation axis.
Finally, to convert the computed quantities to their dimensional
counterparts, we must multiply velocities, densities, and magnetic
fields by $R_X\Omega_X$, $\dot M_w/4\pi R_X^3\Omega_X$, and
$(\Omega_X\dot M_w/R_X)^{1/2}$, respectively.

For interpolations or extrapolations in $\bar \beta$, we recommend
computation first of the dimensionless density, velocity, and
magnetic fields for the three cases $\bar \beta =$ 1, 2, 3, and then
direct interpolations or extrapolations of those fields.  Other
techniques starting farther back in the process run the danger of
obtaining complex roots of $\cal A$ (i.e., complex values of $\rho$)
from the solution of the quartic equation (\ref{Bernoulli}) because
of slight inaccuracies in computing the numerical coefficients.

\subsection{Summary}

In this paper, we have presented a technique by which solutions to
the so-called Grad-Shafranov equation for X-wind flow can be solved,
not by attacking the partial differential equation directly, but by
choosing trial functions that minimize an appropriate action
integral.  While this method has been applied before in problems of
plasma confinement in the fusion community, we believe that the
example given here is its first application in astrophysics for the
notorious case when magnetohydrodynamical flows cross critical
surfaces that change the character of the underlying PDE.

Many empirical arguments suggest that funnel flows and X-winds do
underlie the accretion hot-spots, jets, and winds of YSOs, although
a dipolar field geometry near the star (see Fig. 1) may be an
over-simplification (Ardila et al. 2002, Unruh et al. 2004,
Johns-Krull 2007).  Fortunately, although the fractional areal
coverage of hot spots depends on the detailed multipole structure of
the surfaces of actual young stars, the general validity of X-wind
theory depends only on the level of trapped flux in the X-region and
is insensitive to the magnetic geometry on the star as long as the
fields are strong (Mohanty \& Shu 2007).  The trapped flux in the
X-wind models of this paper are computed as
\begin{equation}
2\pi \bar\beta \left(GM_*\dot M_w\over \Omega_X\right)^{1/2},
\end{equation}
and should be compared with the magnetic flux (area times mean
field) in hot-spots on one hemisphere's surface of the star impacted
by the corresponding funnel flow.  (Both fluxes are 1/3 of the total
trapped flux in the X-region and equal the net flux of the dead
region.)  For T Tauri stars, the comparison is pretty good (see,
e.g., Johns-Krull \& Gafford 2002).

Apart from relative simplicity, the semi-analytical solutions
summarized in \S 7.1 have many other advantages.  For example, the
solutions hold over a formally infinite dynamic range, showing the
asymptotic, logarithmically slow, collimation into jets missing in
many numerical simulations.  These properties make the models of
this paper especially suitable for a wide variety of astronomical
and meteoritical applications, such as detailed comparisons with
observations, trajectories of solids entrained in the wind, and
interactions with neighboring circumstellar or interstellar matter.
A needed generalization for future research is the inclusion of the
effects of the intrinsic magnetization of the surrounding accretion
disk.

\bigskip\bigskip
We thank the Physics Department and the Center for Astrophysics and
Space Sciences of UCSD for support. The Academia Sinica and the
National Science Council (NSC) of Taiwan also provided funding
through their grants to the Theoretical Institute for Advanced
Research in Astrophysics (TIARA). The research of MJC is supported
in part by the NSC grant 95-2112-M-001-44.

\appendix \section{Character of Governing Equation}\label{Character}

The GSE \eqref{GSE-general} resembles the steady state heat
diffusion equation with a variable diffusion coefficient $\A$.  This
analogy is actually misleading since we do not know its overall
character until we substitute in the implicit dependence of $\A$ on
$\psi$ by solving the (algebraic) BE and examine the characteristics
of the GSE.  To do so, let us first differentiate the BE with
respect to $\varpi$ and $z$.
\begin{eqnarray*}
  &&2(\psi_{,\varpi} \psi_{,\varpi \varpi} + \psi_{,z}\psi_{,z \varpi}) -
  \frac{2\A_{,\varpi}}{\A^3}\prn{\frac{J}{\varpi^2} -1}^2 \\
  &&\qquad +
  \frac{2\varpi^4 \A_{,\varpi}}{(\oneoverrho)^3}\brk{2\Veff +
  2\epsilon^2 \ln \prn{\frac{\epsilon^2 h}{\oneoverrho}}+ \epsilon^2}
  + ... = 0,\\
  &&2(\psi_{,\varpi} \psi_{,\varpi z} + \psi_{,z}\psi_{,z z}) -
  \frac{2\A_{,z}}{\A^3}\prn{\frac{J}{\varpi^2} -1}^2 \\
  &&\qquad +
  \frac{2\varpi^4 \A_{,z}}{(\oneoverrho)^3}\brk{2\Veff +
  2\epsilon^2 \ln \prn{\frac{\epsilon^2 h}{\oneoverrho}}+ \epsilon^2}
  + ... = 0,
\end{eqnarray*}
where in the above equations, a subscript denotes partial derivative
and the ellipsis symbols include terms that are irrelevant in
determining the character of the GSE. These equations may be solved
for $\A_{,\varpi}$ and $\A_{,z}$ to give
\begin{displaymath}
  \A_{,\varpi} = \oneover{\mathcal{P}}\prn{\psi_{,\varpi} \psi_{,\varpi\varpi} +
  \psi_{,z} \psi_{,z\varpi}} + ..., \qquad \A_{,z} = \oneover{\mathcal{P}}
  \prn{\psi_{,\varpi} \psi_{,\varpi z} + \psi_{,z}\psi_{,zz}} + ...,
\end{displaymath}
where
\begin{displaymath}
  \mathcal{P} =
  \frac{\varpi^2}{\oneoverrho}
  \abs{\Del \psi}^2 + \oneover{\A^3}\prn{\frac{J}{\varpi^2} -1 }^2 \frac{\beta^2}{\oneoverrho}
  - \frac{\epsilon^2\varpi^4}{(\oneoverrho)^3},
\end{displaymath}
after we eliminate $\Veff$ in the expression by using the BE
\eqref{BE}.  The second derivative terms in the GSE \eqref{GSE} can
now be written in the form
\begin{displaymath}
  a \psi_{,\varpi\varpi} + 2b\psi_{,\varpi z} + c \psi_{,zz} + ... =
  0,
\end{displaymath}
where
\begin{displaymath}
  a = \A + \frac{\psi_{,\varpi}^2}{\mathcal{P}}, \qquad b =
  \frac{\psi_{,\varpi} \psi_{,z}}{\mathcal{P}}, \qquad c = \A +
  \frac{\psi_{,z}^2}{\mathcal{P}}.
\end{displaymath}
The character of the GSE is determined by the quantity $\Delta = b^2
- ac$ (Garabedian 1986): it is elliptic, parabolic, or
hyperbolic if $\Delta$ is negative, zero, or positive.  We may
compute $\Delta$ for our GSE explicitly.
\begin{equation}
  \Delta = - \A^2\left\{
  {|\Del \psi|^2 + (J\varpi^{-2} -1 )^2\A^{-2}
  - \epsilon^2\A \varpi^4[\beta (\beta^2-\varpi^2\A)]^{-2}\over
  \varpi^2 \A\beta^{-2} |\Del \psi|^2 + (J\varpi^{-2} -1)^2\A^{-2}
  - \epsilon^2 \A \varpi^4[\beta (\beta^2-\varpi^2\A)]^{-2}}\right\}.
\end{equation}
The interpretation of this expression becomes transparent if we
transform back into the physical quantities.  After some algebra, we
have
\begin{equation}
  \Delta = - \A^2\brk{
  \frac{u^2
  - \epsilon^2(1-M_A^2)}{
  (1-M_A^2) (u_p^2-\epsilon^2) + u_\varphi^2}}
  =\A^2\brk{
  \frac{(v_A^2 + \epsilon^2) (u_p^2
  - v_s^2)}{(u_p^2 - v_{-p}^2)(u_p^2 - v_{+p}^2)
  }}.\label{PDE_character}
\end{equation}
where $v_{Ap} \equiv \sqrt{B_p^2/\rho}$ is the poloidal component of
the Alfv\'en velocity $v_A \equiv \sqrt{B^2/\rho}$ with $B^2 =
B_p^2+B_\varphi^2$, $u_p$ is the poloidal fluid velocity, and $v_s$
is defined by
\begin{displaymath}
  v_s^2 \equiv \frac{\epsilon^2v_{Ap}^2}{v_A^2 + \epsilon^2}.
\end{displaymath}
In the limit where $v_A^2 \gg \epsilon^2$, it reduces to the thermal
sound speed.  In addition, $v_{\pm p}$ denote the poloidal component
of the fast and slow MHD wave speeds, respective, and are given by
\begin{equation}
  v_{\pm p}^2 = \ahalf (v_A^2 + \epsilon^2)\brk{1 \pm \sqrt{1- \frac{4v_s^2}
  {v_A^2 + \epsilon^2}}}.
\end{equation}
A moment of thought reveals that $v_s < v_{-p} < v_{+p}$.  The
significance of the equation \eqref{PDE_character} is now clear. The
governing GSE is elliptic when $u_p^2 < v_s^2$ or $v_{-p}^2 < u_p^2
< v_{+p}^2$, and it is hyperbolic when $v_s^2 < u_p^2 < v_{-p}^2$ or
$u_p^2 > v_{+p}^2$ \citep{Heinemann:1978,Sakurai:1985}.

To be definite, we shall refer to the loci where the poloidal
velocity squared equals $v_s^2$, $v_{-p}^2$ and $v_{+p}^2$ as
sonic, slow, and fast surfaces, respectively. Despite
the deceiving appearance of the GSE \eqref{GSE-general}, note that it does not
change character on the Alfv\'en surface when $\A = 0$ (or
equivalently when $M_A = 1$ and the total fluid velocity in the corotating frame
equals the total Alfv\'en speed); it remains elliptic until the fast
surface. The fact that the asymptotic flow is described by a
hyperbolic PDE is consistent with our physical intuition.  When the
fluid speed is super-magnetosonic, no information can be sent
upstream into the flow.  Thus the asymptotic behavior of the X-wind is determined
by the ``initial condition'' at the place when the fluid velocity
first becomes equal to the fastest signal propagation speed -- a
defining feature of hyperbolic problems.

In the cold limit the discriminant $\Delta$ has the simplification,
\begin{equation}
  \Delta =  -\A^2\left\{
  {|\Del \psi|^2 + (J\varpi^{-2} -1 )^2\A^{-2} \over
  \varpi^2 \A\beta^{-2} |\Del \psi|^2 + (J\varpi^{-2} -1)^2\A^{-2}}\right\}
  = -\A^2\left(\frac{1
  }{1 -u_p^2/v_A^2}\right).
\end{equation}
This equation explicitly states that the transition to the
hyperbolic portion of the solution is done through the fast surface,
where the poloidal fluid velocity is equal to the magnetosonic
speed, which is the total Alfv\'en speed $v_A = B/\sqrt{\rho}$ when
$\epsilon$ is set to zero.  The axial symmetry of the assumed
problem guarantees that any compressions or rarefactions occur only
in the meridional plane, so the relevant speed of signal propagation
in the limit $\epsilon \rightarrow 0$ is the magnetosonic speed
relative to the poloidal motion of the fluid.

\end{document}